\documentclass[twocolumn]{svjour3}          % twocolumn
\smartqed  % flush right qed marks, e.g. at end of proof
\usepackage{graphicx}
\usepackage{euscript,amsmath,amssymb,amsfonts,graphicx,fullpage,bm,color,cite,paralist}

\usepackage{natbib}

\numberwithin{equation}{section}

\usepackage{epstopdf} 

\newcommand{\e}{{\rm e}}
\renewcommand{\d}{{\rm d}}
\newcommand{\pd}{\partial}

\newcommand{\D}{\displaystyle}
\newcommand{\mc}{\mathcal }
\newcommand{\ve}{\varepsilon}

\newcommand{\A}{\mathcal{A}}

\newcommand{\ep}{\epsilon}

\newcommand{\bZ}{\mathbf{Z}}

\begin{document}

\allowdisplaybreaks

\title{Synaptic efficacy shapes resource limitations in working memory}

%\title{Item limitations in working memory modeled by interacting bumps}

\titlerunning{Synaptic efficacy shapes resource limitations in WM}        % if too long for running head

\author{Nikhil Krishnan \and Daniel B. Poll \and Zachary P. Kilpatrick}

\institute{N. Krishnan \at
              Department of Applied Mathematics, University of Colorado Boulder \\
              Boulder CO 80309 USA \\
              \email{Nikhil.Krishnan@colorado.edu} 
              \and
              D.B. Poll \at
              Department of Engineering Sciences and Applied Mathematics, Northwestern University \\
              Evanston IL 60208 USA \\
              \email{daniel.poll@northwestern.edu}
              \and
              Z.P. Kilpatrick \at
              Department of Applied Mathematics, University of Colorado Boulder \\
              Boulder CO 80309 USA \\
              \email{zpkilpat@colorado.edu}    
}

\date{Received: date / Accepted: date}

\maketitle

\begin{abstract}

Working memory (WM) is limited in its temporal length and capacity. Classic conceptions of WM capacity assume the system possesses a finite number of slots, but recent evidence suggests WM may be a continuous resource. Resource models typically assume there is no hard upper bound on the number of items that can be stored, but WM fidelity decreases with the number of items. We analyze a neural field model of multi-item WM that associates each item with the location of a bump in a finite spatial domain, considering items that span a one-dimensional continuous feature space. Our analysis relates the neural architecture of the network to accumulated errors and capacity limitations arising during the delay period of a multi-item WM task. Networks with stronger synapses support wider bumps that interact more, whereas networks with weaker synapses support narrower bumps that are more susceptible to noise perturbations. There is an optimal synaptic strength that both limits bump interaction events and the effects of noise perturbations. This optimum shifts to weaker synapses as the number of items stored in the network is increased. Our model not only provides a neural circuit explanation for WM capacity, but also speaks to how capacity relates to the arrangement of stored items in a feature space.

% there is controversy about whether working memory capacity is accounted for by slots or continuous resources
% many of these models tend to be non-mechanistic fits to psychophysical data
% recently some authors have proposed models involving interacting bumps
% thorough analysis would help to link features of the network connectivity to capacity and the timescale of error
% we use interface dynamics, which provides a low-dimensional projection of the dynamics
% findings about how network connectivity relates to capacity
% do some performance plots

\keywords{bump attractor, working memory, capacity, limited resource, interface methods}
\end{abstract}

% Section 1
\section{Introduction}
\label{intro}

Working memory (WM) is defined by both its short timescale and its capacity limitations~\citep{ma14}. Detailed behavioral and electrophysiological recordings have demonstrated that WM is associated with persistent neural activity in a number of cortical regions~\citep{constantinidis16}. Neural and synaptic activity fluctuations account for commonly observed errors accumulated during the delay-period of typical WM tasks~\citep{compte00,wimmer14}. However, there is controversy surrounding the origin of errors arising from limitations of WM capacity. Classic models contend that item-limits are best defined by a `slot model,' placing a hard upper bound on the number of items that can be stored~\citep{luck97,cowan10}. On the other hand, recent evidences suggests a `resource model,' with no hard item-number limit, in which a fixed continuous resource is spread across an arbitrary number of items to be remembered~\citep{wilken04,bays08,vandenberg12,keshvari13}.

Both the slots model and the resource model reproduce some gross statistics from WM tasks equally well. For example, the recall variability tends to increase in both models for a WM task with a high number of items~\citep{luck13,bays09}. However, for lower item counts, response variability is flat in a slots model whereas it increases in a resource model. In addition, more task-relevant cues can be stored with higher precision in a continuous resource model, but not in a slots model. Recent experiments have demonstrated that WM precision varies for low item counts~\citep{bays09} and for cued versus uncued items~\citep{gorgoraptis11}. These results have been recapitulated in several human and non-human primate studies, suggesting flexibility in the allocation of WM~\citep{buschman11,lara12}. Resource models of WM allow for such flexibility, suggesting many possibilities for how storage precision varies across task parameters~\citep{vandenberg12,fougnie12}.

\begin{figure}
\begin{center} \includegraphics[width=7.6cm]{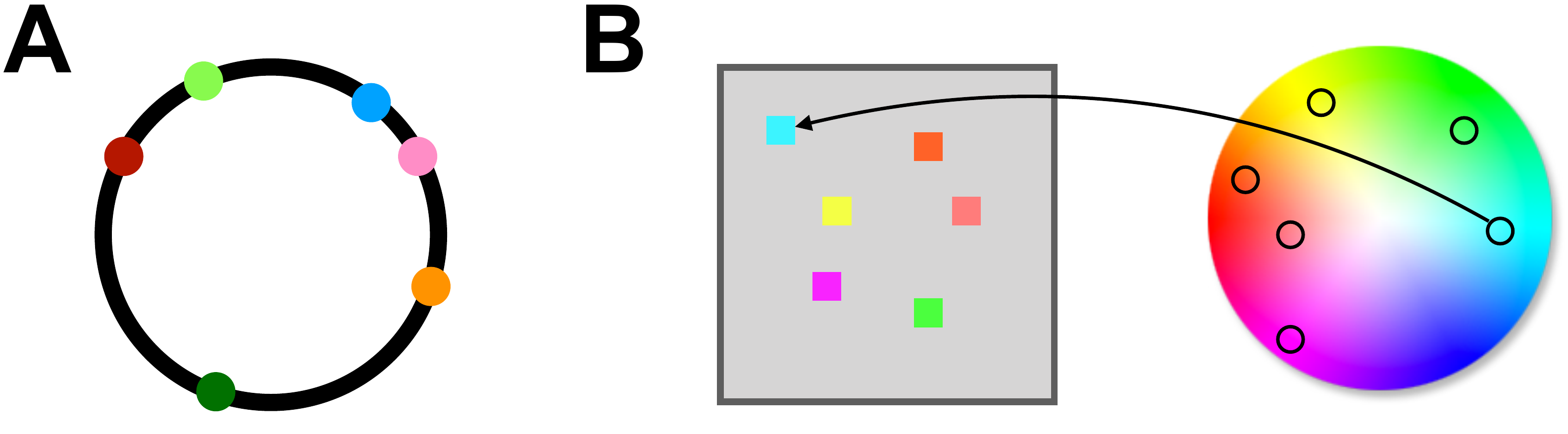} \end{center}
\vspace{-2mm}
\caption{Examples of multi-item visual stimuli used in working memory (WM) tasks~\citep{zhang08,bays09,ma14}. {\bf A}. Memoranda here are angles on a circle corresponding to the dot locations, identified by their color. Subject will be required to memorize all objects and then just recall one item; e.g., the location of the blue dot. {\bf B}.~Alternatively, subjects may have to memorize the color of each item. For example, a subject may be asked what the color of the top left square was.}
\vspace{-4mm}
\label{fig1_expex}
\end{figure}

Computational models that capture behavioral patterns in multi-item WM are an active area of research~\citep{barak14}. It remains an open question what neural mechanisms underlie these trends in response variability. Recent studies have extended the framework of continuous attractor networks, successful in capturing error accumulation in single-item WM tasks~\citep{wimmer14}, to account for errors observed in multi-item WM~\citep{edin09,wei12,almeida15}. These models are well-suited to store memoranda drawn from a continuous space, such as locations and colors (See Fig. \ref{fig1_expex}). Recurrent networks comprised of a locally excitatory population coupled to a broadly tuned inhibitory population produce ``bumps" of persistent neural activity~\citep{amari77,compte00}. Bumps encode the remembered location of a presented angle during the WM delay period, and fluctuations arising from stochastic spiking or synaptic transmission degrade memory of the initial position~\citep{compte00,kilpatrick13b}. Multi-item WM errors arise in these models via the interactions of multiple bumps, each bump encoding a distinct angle~\citep{edin09,wei12,almeida15}. Bumps can repel, merge, or annihilate one another via nonlocal synaptic interactions of the network. For randomly chosen angles, the relative precision of recall decreases with set size according to a power law~\citep{wei12}, as in \citet{bays08}. Thus, a multiple bumps model of WM appears to reconcile observed behavioral trends with known neural circuit mechanisms for storing WM using persistent activity.

These previous studies were performed using large-scale spiking simulations, however, and could not draw clear connections between parameters and the model's WM performance. An advantage of using neural field equations to describe large-scale network interactions is that they are analytically tractable, and their dynamics can often be approximated by low-dimensional systems that solely describe variables of interest~\citep{bressloff12}. For instance, previous neural field studies of bump attractor models of single-item WM have developed explicit expressions for the relationship between network connectivity and the response variability~\citep{kilpatrick13c,carroll14,kilpatrick17}. Other work has explored the interaction of multiple bumps in neural field equations, but in special cases which are not relevant to the problem of storing an arbitrary set of memoranda~\citep{laing03,bressloff05,laing02,lu11}.
%Construction of stationary bump solutions in a homogenenoulateral inhibitory neural field tends to only work for single bump solutions~\citep{amari77}.
A robust model for storing multiple items would allow for multiple bumps to be stored at arbitrary locations around a network. Our study explores tradeoffs in the strength of neural architecture as it impacts bumps' response to fluctuations, as well as interactions between neighboring bumps.

We utilize interface methods, originally applied to single bump solutions~\citep{amari77,coombes12}, to project the dynamics of multiple bumps in a neural field to a low-dimensional system of differential equations for the edges of the bumps. This approximate system can be analyzed in order to uncover the relationship between the architecture of the network and the robustness of multi-item WM. In particular, we examine how bumps interact with one another, and how they respond to external fluctuations that model the known stochastic evolution of persistent activity during the WM delay-period~\citep{wimmer14}. Interestingly, increasing the strength of synaptic coupling makes networks with bumps that are robust to noise, but at the cost of producing stronger interactions between bumps. As a result, networks with the lowest response variability have an intermediate value of synaptic strength, which trades off the robustness of wide bumps to noise with the increased precision of networks containing narrower bumps.
%Explicit expressions can be derived using our interface equations, and these results compare favorably to numerical simulations of the full neural field model.

% background on multi-item working memory Compte model as well as XJ Wang model
% data from Bays lab on merging of memories. what formula do errors follow?
% interesting feature of Compte model is the merging of bumps can explain memory loss.
% background on bump literature. little work on multiple bumps interacting in a layer. Laing and Troy (2003); Bressloff (2005); Laing et al (2001); Coombes et al (2014). No working low-dimensional approximations for interacting interfaces in a simple Amari neural field
% issue is there are no exact double bump solutions for typical lateral inhibition, but there are near-solutions that drift apart very slowly.

% Section 2
\section{Neural field model of visuospatial working memory}
\label{model}

Most bump attractor models of working memory (WM) focus on tasks where a subject must remember a single orientation each trial~\citep{kilpatrick13,wimmer14}. However, WM capacity can be probed by testing subjects' ability to recall multiple items~\citep{bays08,zhang08}. We analyze a recurrent, spatially-organized network, which can represent multiple orientations during the delay period of a visuospatial WM task~\citep{almeida15}. The model is similar to the case of single-item WM, but the network architecture plays an important role in shaping memory capacity~\citep{bays15}.

\subsection{Model definition}

We study a neural field model where locations of neurons correspond to their preferred stimulus orientation, organized in a ring architecture with slow local excitation and broad inhibition~\citep{ermentrout98}:
\begin{align}
\d u(x,t) &= \left[ - u(x,t) + w(x)*H(u(x,t) - \theta) \right] \d t  \label{nfield} \\
& \hspace{18mm} + \sqrt{\ep \cdot |u(x,t)|} \hspace{0.5mm} \d Z(x,t). \nonumber
\end{align}
The variable $u(x,t)$ represents synaptic input to spatial location $x \in [-L,L]$ at time $t$, which is periodic so $u(L,t) = u(-L,t)$. The weight function $w(x-y)$ represents the synaptic connectivity from neurons at location $y$ to location $x$ via the convolution $w*H(u-\theta) = \int_{- L}^L w(x-y) H(u(y,t)-\theta) \d y$. Note, we assume the weight function is even $w(x) = w(-x)$ and satisfies periodicity $w(-L) = w(L)$.

\begin{figure}
\begin{center} \includegraphics[width=6cm]{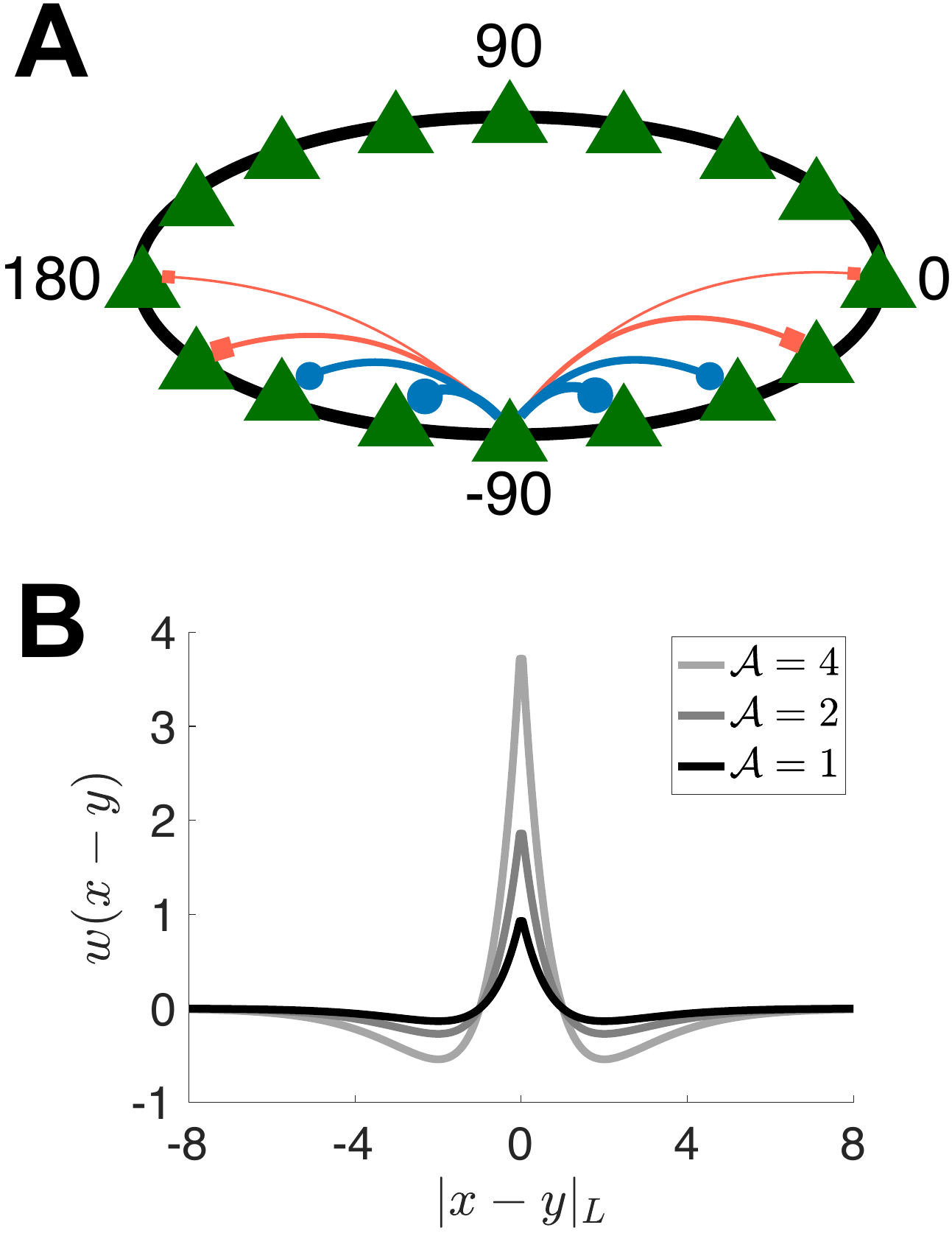} \end{center}
\vspace{-2mm}
\caption{Recurrent network model of multiple-item WM. {\bf A.} The neural field, Eq.~(\ref{nfield}) is comprised of local populations (green triangles) organized on a ring with distance-dependent connectivity. This single layer describes the activity $u(x,t)$ of a combined excitatory/inhibitory neural field, derived in the limit of fast inhibition~\citep{amari77,pinto01b,carroll14}. Strong effective excitation (blue dots) is narrow whereas weaker effective inhibition (red squares) is wide. {\bf B.} Weight function, Eq.~(\ref{wizard}), scaled by different maximal synaptic strengths $\A = {\rm max}_{x \in [-L,L]} w(x)$.}
\vspace{-2mm}
\label{fig2_weight}
\end{figure}

We consider the weight function,
\begin{align}
w(x-y) = {\mc A} (1- |x-y|_L) \e^{-|x-y|_L}, \label{wizard}
\end{align}
with local excitation and broad inhibition~\citep{coombes05}, where $\A$ parameterizes the synaptic strength, and $|x|_L = {\rm min}(|x-y|,|2L - |x-y||)$ is the distance on the ring (Fig. \ref{fig2_weight}). Note, neurons with similar orientation will tend to activate one another, while neurons with dissimilar orientation tend to inhibit one another. Integrating $\int_{-L}^L w(x-y) \d y = 2 \A(1 - \e^{- L}) - 2{\mc A}(1 - (1+L) \e^{-L}) = 2\A L \e^{-L}$, we find the total excitation and inhibition is approximately balanced ($\int_{-L}^L w(x-y) \d y \approx 0$) for $L \gg 1$. Since we are interested in focusing on how scaling $\A$ impacts WM capacity for angles, we fix $L : = 180$ consistent with typical oculomotor delayed-response tasks~\citep{funahashi89,goldmanrakic95,wimmer14,constantinidis16}.
%This is in the limit of large domains $L \gg 1$, so we take $\e^{-L} \approx 1$ for most of our calculations, which simplifies formulas considerably. We therefore take $\A_e = \A_i = \A$, yielding
%\begin{align}
%w(x-y) = \A(1 - |x-y|_L) \e^{-|x-y|_L},  \label{wizard}
%\end{align}
%where $\A$ controls the strength of both excitation and inhibition, which are balanced so $\int_{-L}^L w(x-y) \d y \equiv 0$.
%Subsequently, we examine the impact of the synaptic strength $\A$ on the limitations of WM in Eq.~(\ref{nfield}).
The weight function Eq.~(\ref{wizard}) is one example in a class of synaptic kernels that arises in the limit of fast inhibition, such that separate excitatory and inhibitory populations can be combined into a single population Eq.~(\ref{nfield})~\citep{carroll14}. The nonlinearity in Eq.~(\ref{nfield}) is a Heaviside step function
\begin{align}
H(u - \theta) &= \left\{ \begin{array}{cc} 1, & u> \theta, \\ 0, & u< \theta, \end{array} \right.  \label{H}
\end{align}
representing the input-output relationship between synaptic activation and population firing rate at $x$. Smooth sigmoids are also often used, but the qualitative dynamics of Eq.~(\ref{nfield}) remain similar for steep sigmoids~\citep{coombes10,bressloff12}.
%See \citet{kilpatrick10b} for issues that arise when when auxiliary variables (e.g., adaptation or synaptic depression) have an exposed discontinuity in their evolution equations.
We exploit the fact that the output of $H(u-\theta)$ is binary ($\{0,1\}$) to develop interface equations for the dynamics of bumps in Eq.~(\ref{nfield}), adapting methods originally developed by \citet{amari77} and extended by \citet{coombes12}.

Rather than modeling instantiation of bumps in Eq.~(\ref{nfield}) using a spatiotemporal input as in \citet{compte00,kilpatrick13b,almeida15}, we consider bump initiation implemented with initial conditions. Initiating bumps with external input does not significantly alter our results. Stochasticity is modeled by weak and multiplicative noise $\sqrt{\ep \cdot |u(x,t)|} \d Z(x,t)$, driven by the increment of a spatially-dependent Wiener process such that $\langle \d Z(x,t) \rangle = 0$ and $\langle \d Z(x,t) \d Z(y,s) \rangle = C(x-y) \delta (t-s) \d t \d s$. The spatial correlations $C(x-y)$ are a symmetric function that depends on the distance between two locations in the network. Typical formulations of Langevin equations take the multiplicative noise to be of Stratonovich form~\citep{gardiner09}, and a related neural field equation can be derived by applying a Kramers-Moyal expansion to a neural master equation~\citep{bressloff09}.  

The impact of multiplicative noise on bump dynamics is analyzed by adapting methods developed for bumps in neural fields with additive noise~\citep{kilpatrick13}. Note, we could modify Eq.~(\ref{nfield}) to account for the systematic shift induced by multiplicative noise in the Stratonovich sense. However, this contribution will be ${\mc O}(\ep)$ in comparison to the ${\mc O}(\sqrt{\ep})$ amplitude of the noise itself. Thus, we simply truncate the equation to ignore these additional terms, which would only slightly shift the resulting form of the stationary solution we will linearize about, as discussed in \citet{bressloff12b}. To be explicit, we note that we can compute $\sqrt{\ep} \langle u(x,t) \d Z(x,t) \rangle = \ep C(0) \langle {\rm sign} [u(x,t)] \rangle \d t/2$~\citep{novikov65}, smaller than the $\sqrt{\ep}$-amplitude noise term we consider. Note, we have run simulations of both the original Eq.~(\ref{nfield}) and the associated mean-corrected equations, and the results are not noticeably different.
%\begin{align}
%\d u(x,t) &= \left[ g(u(x,t)) + \int_{-L}^{L} w(x-y) H(u(y,t) - \theta) \d y  \right] \d t + \sqrt{\ep} \d R(u,x,t), \label{nfieldcor}
%\end{align}
%where
%\begin{align*}
%g(u(x,t)) = -u(x,t) + \frac{\ep C(0) {\rm sign}[u(x,t)]}{2}
%\end{align*}
%and
%\begin{align*}
%\d R(u,x,t) = \sqrt{|u(x,t)|} \d Z(x,t) - \frac{\sqrt{\ep} C(0) {\rm sign}[u(x,t)]}{2} \d t,
%\end{align*}
%so the resulting stochastic process $R$ has zero mean and variance
%\begin{align*}
%\langle \d R(u,x,t) \d R(u,y,t) \rangle = \langle \sqrt{|u(x,t)|} \d Z(x,t) \sqrt{|u(y,s)|} \d Z(y,t) \rangle + {\mc O}(\sqrt{\ep}).
%\end{align*}
%We will use the modified Eq.~(\ref{nfieldcor}) in order to study the interactions of bumps in the stochastic version of the neural field Eq.~(\ref{nfield}). First, we review the existence of single stationary bump solutions in the noise-free ($Z\equiv 0$) version of Eq.~(\ref{nfield}).

\subsection{Single bump solutions}

Solutions to the noise-free ($Z \equiv 0$) version of Eq.~(\ref{nfield}) can be found explicitly for specific weight functions~\citep{bressloff12}. In particular, single bump (stationary pulse) solutions exist when $w(x)$ satisfies requirements making it laterally inhibitory~\citep{amari77}, as Eq.~(\ref{wizard}) is. We construct this solution and demonstrate the use of the interface method for characterizing non-equilibrium dynamics of perturbed bump solutions. This will guide our understanding for applying the interface method to multiple bumps.

In the absence of stochasticity ($Z \equiv 0$), stationary solutions to Eq.~(\ref{nfield}) satisfy $u(x,t) \equiv U(x)$, leading to the implicit equation
\begin{align}
U(x) = \int_{-L}^L w(x-y) H(U(y)-\theta) \d y.   \label{stateq}
\end{align}
Unimodal stationary bumps possess a simply-connected active region $\bar{A} = [\bar{x}_1, \bar{x}_2] = \{ x| U(x) \geq \theta \} $ (assuming $-L \leq \bar{x}_1 < \bar{x}_2 < L$), which allows us to rewrite Eq.~(\ref{stateq}) as
\begin{align}
U(x) = \int_{\bar{x}_1}^{\bar{x}_2} w(x-y) \d y.  \label{bumpint}
\end{align}
For analytical convenience, the translation symmetry of the network Eq.~(\ref{nfield}) can be utilized to shift solutions $U_0(x) = U(x - (\bar{x}_1+\bar{x}_2)/2)$ so they are centered at zero~\citep{ermentrout98,bressloff12}:
\begin{align}
U_0(x) &= \int_{-h}^h w(x-y) \d y \label{bumpcent} \\
&= W(x+h)-W(x-h), \nonumber
\end{align}
where $h = (\bar{x}_2 - \bar{x}_1)/2$ and we have defined the antiderivative
\begin{align}
W(x) = \int_0^x w(y) \d y. \label{wanti}
\end{align}
%We show this by using Eq.~(\ref{bumpint}), and writing
%\begin{align*}
%U(x - (\bar{x}_1+\bar{x}_2)/2)
%%&= \int_{\bar{x}_1}^{\bar{x}_2} w(x - (\bar{x}_1+\bar{x}_2)/2 -y) \d y \\
%& = \int_{ - (\bar{x}_2 - \bar{x}_1)/2}^{(\bar{x}_2 - \bar{x}_1)/2} w(x-z) \d z = U_0(x),
%\end{align*}
%where we made the substitution $z = y + (\bar{x}_1+\bar{x}_2)/2$.

Linear stability of stationary bumps can be determined by examining the evolution of perturbations $u(x,t) = U_0(x) + \ve \psi (x,t) + {\mc O}(\ve^2)$ to the bump profile. Linearizing Eq.~(\ref{nfield}) leads to the following evolution equation for the perturbation
\begin{align}
\psi_t(x,t) = - \psi (x,t) + w*\left[H'(U_0 - \theta) \psi\right].  \label{linstab}
\end{align}
Separability of solutions $\psi (x,t) =  \e^{\lambda t} \bar{\psi} (x)$ can be shown~\citep{carroll14}, yielding
%, and for Heaviside nonlinearities it is possible to project the resulting eigenvalue problem to a low-dimensional equation for the evolution of the boundaries of the bump~\citep{amari77,pinto01b}. We focus on
the integral equation for linear stability
\begin{align}
(\lambda + 1) \bar{\psi} (x) = w(x)*\left[H'(U_0(x) - \theta) \bar{\psi} (x) \right].  \label{eigprob}
\end{align} 
Note, $(\lambda, \bar{\psi} (x)) = (0, U_0'(x))$ is a solution, since by plugging in we find
\begin{align}
U_0'(x) = w(x)*\left[H'(u(x,t) - \theta) U_0' (x)\right],  \label{uder}
\end{align}
and Eq.~(\ref{uder}) arises by differentiating the stationary bump Eq.~(\ref{stateq}). This indicates the bump is marginally stable to perturbations that shift its position~\citep{amari77,kilpatrick13}, the main source of error when considering noise and interactions with other bumps.

\begin{figure}
\begin{center} \includegraphics[width=6cm]{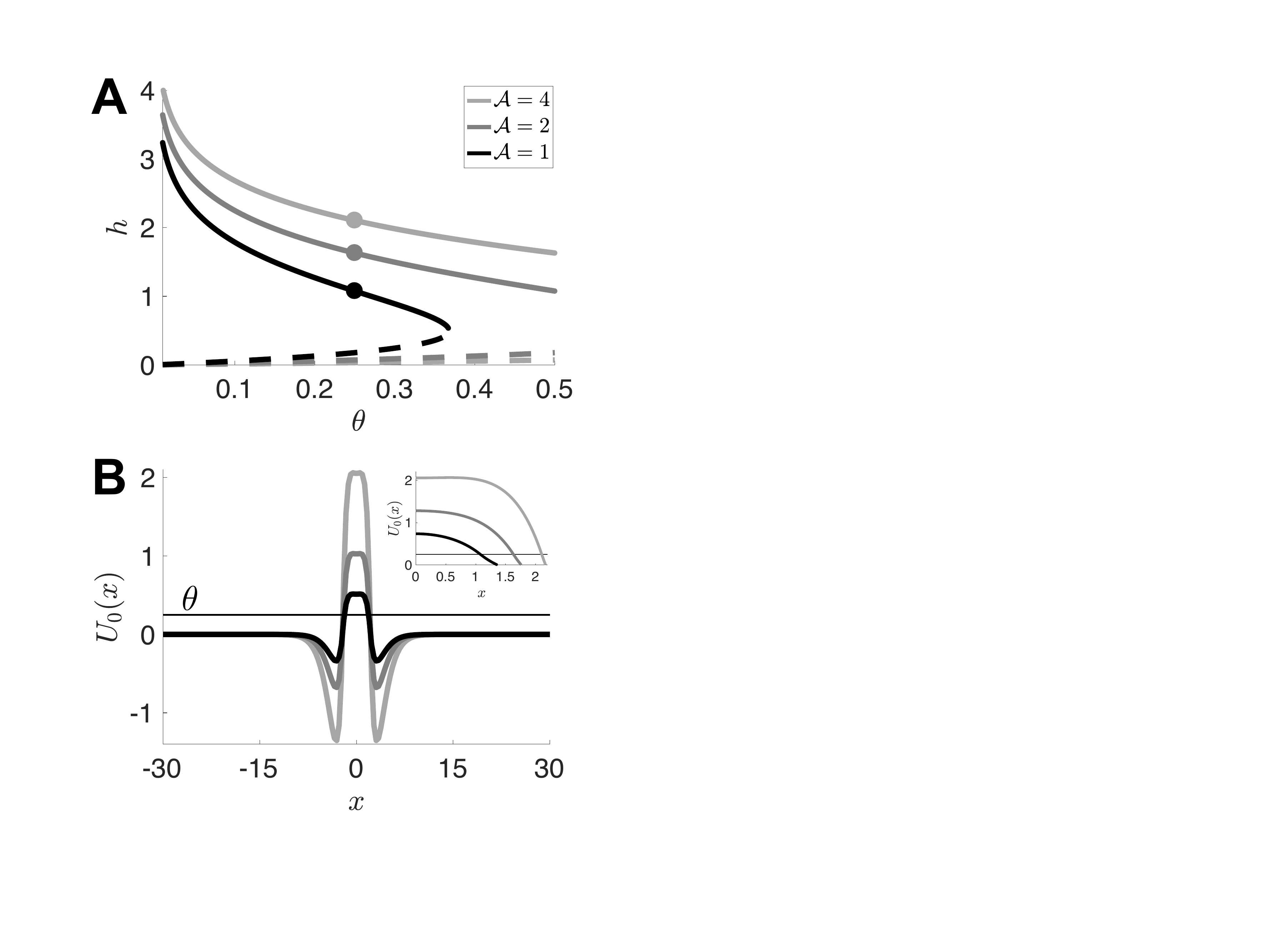} \end{center}
\caption{{\bf A.} Width of bumps, stable (solid) and unstable (dashed), computed using the threshold condition, Eq.~(\ref{widform}). Stable bump width increases with the synaptic strength $\A$. {\bf B.} Examples of bump profiles corresponding to the dots in {\bf A}, widening as $\A$ is increased for fixed $\theta = 0.25$. Inset shows zoom-in of right threshold crossings.}
\label{fig3_bump}
\end{figure}

We can explicitly construct the eigensolutions to Eq.~(\ref{eigprob}) by applying the identity $H'(U_0(x)-\theta) = \bar{\alpha} \left[ \delta (x-h) + \delta (x+h) \right]$ with $\bar{\alpha} = |U_0'(\pm h)| = w(0) - w(2h)$~\citep{amari77,bressloff12}. We find
\begin{align*}
(\lambda + 1) \bar{\psi} (x) = \bar{\alpha} \left[ w(x-h) \bar{\psi} (h) + w(x+h) \bar{\psi}(-h) \right],
\end{align*}
so solutions only depend on the values of $\bar{\psi}(x)$ at $x = \pm h$. Assuming $\bar{\psi}(h) = - \bar{\psi}(-h)$, we find the associated eigenvalue is $\lambda_o = 0$, demonstrating the bump is marginally stable to odd perturbations as mentioned above. For even perturbations $\psi (h) = \psi (-h)$, we find the associated eigenvalue
\begin{align}
\lambda_e = \frac{2 w(2h)}{w(0) - w(2h)}. \label{eveig}
\end{align}
Thus, the stability of the bump will be determined by the sign of $w(2h)$. Typically, the wider bump has $w(2h)<0$ (Fig. \ref{fig3_bump}A), so it is linearly stable~\citep{amari77,kilpatrick16}.

%There are many canonical weight functions for which Eq.~(\ref{bumpcent}) can be integrated explicitly. As it will be used in our subsequent calculations,
We perform the integral in Eq.~(\ref{bumpcent}) for the case of a weight function of form Eq.~(\ref{wizard}), noting the antiderivative Eq.~(\ref{wanti}) is thus given 
\begin{align}  
W(x)  = \A \int_0^x (1-|y|) \e^{-|y|} \d y = \A x \e^{-|x|}.  \label{wizanti}
\end{align}
Note, for analytical convenience, we approximate $L \to \infty$, obtaining~\citep{coombes05}:
\begin{align}
U_0 (x) = \A \left[ (x + h) \e^{-|x+h|} - (x-h) \e^{-|x-h|} \right].   \label{bumpsol}
\end{align}
Self-consistency requires the threshold conditions $U(\pm h) = \theta$ be satisfied, yielding an implicit equation for the bump half-width
\begin{align}
G(h) : = U_0(\pm h) = W(2h) = 2 \A h \e^{-2h} = \theta.   \label{widform}
\end{align}
We show $h(\theta)$ for different values of synaptic strength ${\mc A}$ in Fig.~\ref{fig3_bump}A, and plot stable bump solutions in Fig.~\ref{fig3_bump}B, showing they expand in width as the synaptic strength $\A$ is increased.

There are both wide stable and narrow unstable bumps of form Eq.~(\ref{bumpsol}). A critical value of $\theta$ defines the point where these branches of Eq.~(\ref{widform}) annihilate in a saddle-node (SN) bifurcation~\citep{kilpatrick16}. Differentiating with respect to $h$, the SN bifurcation occurs where $G'(h) = (1-2h) 2 {\mc A} \e^{-2h} = 0$ which can be solved for $h_c=1/2$. Plugging $h_c = 1/2$ into Eq.~(\ref{widform}), we find $\theta_c = {\mc A} \e^{-1}$. Thus, we select $\theta < \theta_c$ given synaptic strength $\A$ to ensure solution existence. Furthermore, we can differentiate Eq.~(\ref{widform}) with respect to the synaptic strength ${\mc A}$, yielding
\begin{align*}
\frac{\d h}{\d {\mc A}} = \frac{h}{{\mc A}(2h-1)}  > 0,
\end{align*}
for $h>1/2$, which occurs for stable bumps as long as $\theta< \theta_c$. Thus, the width of stable bumps will always increase as ${\mc A}$ is increased. Before exploring interactions of multiple bumps of the form of Eq.~(\ref{bumpsol}), we discuss the interface method we will use to obtain low-dimensional approximations for bump dynamics.

%Bump solutions to the modified Eq.~(\ref{nfieldcor}) can also be identified, in the case for which we set $R \equiv 0$ but keep the $C(0)$-dependent correction term $g(u)$. These solutions are useful in asymptotic calculations of the stochastic motion of noise-driven bumps. The systematic drift induced by multiplicative noise changes the profile of the underlying average bump solution. Setting $R \equiv 0$ in Eq.~(\ref{nfieldcor}) and looking for stationary solutions centered at $x=0$, we can follow similar calculations to the above to yield
%\begin{align}
%U_{\ep}(x) = {\mc A} \left[ (x+h) \e^{-|x+h|} - (x-h) \e^{-|x-h|} \right] + \frac{\ep}{2} {\mc S}(x),  \label{bnosol}
%\end{align}
%where ${\mc S}(x) =1$ when $|x|<h_2$, ${\mc S}(x) = 0$ when $|x|=h_2$, and ${\mc S}(x) = -1$ when $|x|>h_2$, and $U_0(h_2) = 0$, using Eq.~(\ref{bumpsol}). Self-consistency yields an analogous implicit equation for the bump half-width $2 {\mc A}h \e^{-2h} = \theta - \ep C(0)/2 = \theta_{\ep}$. Thus, as $\ep$ is increased, the width of stable bumps increases.
%, related to the finding that wavespeed increases with noise amplitude in excitatory neural field equations~\citep{bressloff12b}.

\subsection{Interface equations for a single bump}

Motivated by solutions of the stationary bump type, as in Eq.~(\ref{bumpint}), \citet{amari77} developed an interface theory for excitation patterns in the noiseless ($Z \equiv 0$) version of the neural field Eq.~(\ref{nfield}). This approach was recently reviewed in \citet{amari14}, and has been extended to capture the dynamics of other solutions by \citet{coombes11,coombes12,avitabile17}. We extend these techniques further to account for stochastic perturbations due to the noise term in Eq.~(\ref{nfield}). Interface equations are derived by noting that the output of the Heaviside nonlinearity, Eq.~(\ref{H}), is determined by the active region $A(t) = \{ x | u(x,t) \geq \theta \}$ of the spatial domain $x \in [-L,L)$. For a single bump, we define the active region $A(t) = [x_1(t), x_2(t)]$, where the interfaces occur at the boundary points $x_1(t)$ and $x_2(t)$,
\begin{align}
u(x_j(t), t) = \theta, \hspace{5mm} j=1,2.  \label{ipoints}
\end{align}
We rewrite Eq.~(\ref{nfield}), using our assumed form of the active region $A(t)$ as
\begin{align}
\d u(x,t) =& \left[ -u(x,t) + \int_{x_1(t)}^{x_2(t)} w(x-y) \d y \right] \d t \nonumber \\
& \hspace{10mm} + \sqrt{\ep \cdot |u(x,t)|} \d Z(x,t).  \label{nfact}
\end{align}
We now derive a stochastic evolution equation for the interfaces, $\d x_j = d_j(x_1,x_2,t) \d t + g_j(x_1,x_2,t) \d z_j$, where $d_j$ is a drift and $g_j$ corresponds to the diffusion term. Differentiating Eq.~(\ref{ipoints}) with respect to time, we obtain the following consistency equation for the location of the interfaces $x_j(t)$ and the evolution of the activity variable
\begin{align}
\alpha_j(t) \d x_j(t) + \beta_j(t) (\d x_j(t))^2 + \d u(x_j(t),t) = 0, \label{ifcon}
\end{align}
for $j = 1,2$, where we have defined the spatial gradient at the interface points
\begin{align*}
\alpha_j(t) = \frac{\pd u(x_j(t),t)}{\pd x}, \hspace{9mm} j = 1,2,
\end{align*}
and the second derivative $\beta_j(t) = \frac{1}{2} u_{xx}(x_j(t),t)$ for $j=1,2$. The middle term in Eq.~(\ref{ifcon}) arises from an application of It\^{o}'s lemma~\citep{gardiner09}. For simplicity, we approximate the spatial gradients using that of the stationary solution for now, $\alpha_1(t) \approx \bar{\alpha} =  U_{0}'(-h)$ and $\alpha_2(t) \approx - \bar{\alpha}= U_{0}'(h) = -U_{0}'(-h)$, computed directly from Eq.~(\ref{bumpsol}). In \citet{coombes12,gokcce17}, the dynamic evolution of the gradients $\alpha_j(t)$ is tracked in the case of a deterministic system ($Z \equiv 0$ in Eq.~(\ref{nfield})). As, discussed, we drop $o (\ep)$ terms resulting from multiplicative noise, for simplicity. Thus, the middle term in Eq.~(\ref{ifcon}) will vanish, since the noise will have amplitude ${\mc O}(\sqrt{\ep})$, as we show, and the other terms in $(\d x_j)^2$ are vanishingly small. Using the evolution equation for the neural activity, Eq.~(\ref{nfact}), and the interface condition Eq.~(\ref{ipoints}), we can describe the evolution of the interfaces by rearranging Eq.~(\ref{ifcon}) to find
\begin{align}
\d x_j(t) &= \frac{(-1)^j}{\bar{\alpha}} \left( \left[ -\theta + \int_{x_1(t)}^{x_2(t)} w(x_j(t)-y) \d y  \right] \d t \right. \nonumber \\
& \hspace{8mm} \left. + \sqrt{\ep \cdot \theta} \hspace{0.5mm} \d Z(x_j(t),t) \right)  \label{ifacea}
\end{align}
for $j=1,2$. Since the integral in Eq.~(\ref{ifacea}) can be evaluated, we employ our definition, Eq.~(\ref{wanti}), of the antiderivative $W(x)$ and write
\begin{align*}
\int_{x_1(t)}^{x_2(t)} w(x_j(t)-y) \d y = W(x_2(t) - x_1(t)),
\end{align*}
yielding an even simpler form for the interface equations
\begin{align}
\d x_j(t) &= \frac{(-1)^j}{\bar{\alpha}} \left( \left[ - \theta + W(x_2(t) - x_1(t)) \right] \d t \right. \nonumber \\
& \hspace{22mm} \left. + \sqrt{\ep \cdot \theta} \hspace{0.5mm} \d Z(x_j(t),t) \right) \label{ifaceb}
\end{align}
for $ j = 1,2$. We now remark on a number of observations to be made concerning Eq.~(\ref{ifaceb}). First, in the absence of noise ($Z \equiv 0$), there is a line of fixed points to the resulting equation
\begin{align}
\frac{\d x_j}{\d t} =  \frac{(-1)^j}{\bar{\alpha}} \left( - \theta + W(x_2(t) - x_1(t))  \right) \label{detone}
\end{align}
for $ j = 1,2$, in the space $(x_1,x_2)$ satisfying $W(x_2 - x_1) = \theta$, which is precisely Eq.~(\ref{widform}) in the case that $x_2 - x_1 = 2h$. Also, note that when the interface locations are symmetric about $x=0$, then $x_2(t) = -x_1(t) = a(t)$ can be described by a single equation by plugging into Eq.~(\ref{detone}) to yield
\begin{align}
\frac{\d a}{\d t} =  \frac{1}{\bar{\alpha}} \left( - \theta + W(2 a(t))  \right).  \label{oneone}
\end{align}
%Second, we note that positive deterministic inputs ($I(x,t) \geq 0$) will tend to expand the active region outward, since if initially Eq.~(\ref{bstat}) is satisfied, and then a nonzero input is applied at some time $t$, then
%\begin{align*}
%\frac{\d x_1}{\d t} = - \frac{I(x_1(t),t) }{\bar{\alpha}} <0, \hspace{5mm} \frac{\d x_2}{\d t} = \frac{I(x_2(t),t)}{\bar{\alpha}}  >0.
%\end{align*}
\begin{figure*}
\begin{center} \includegraphics[width=16cm]{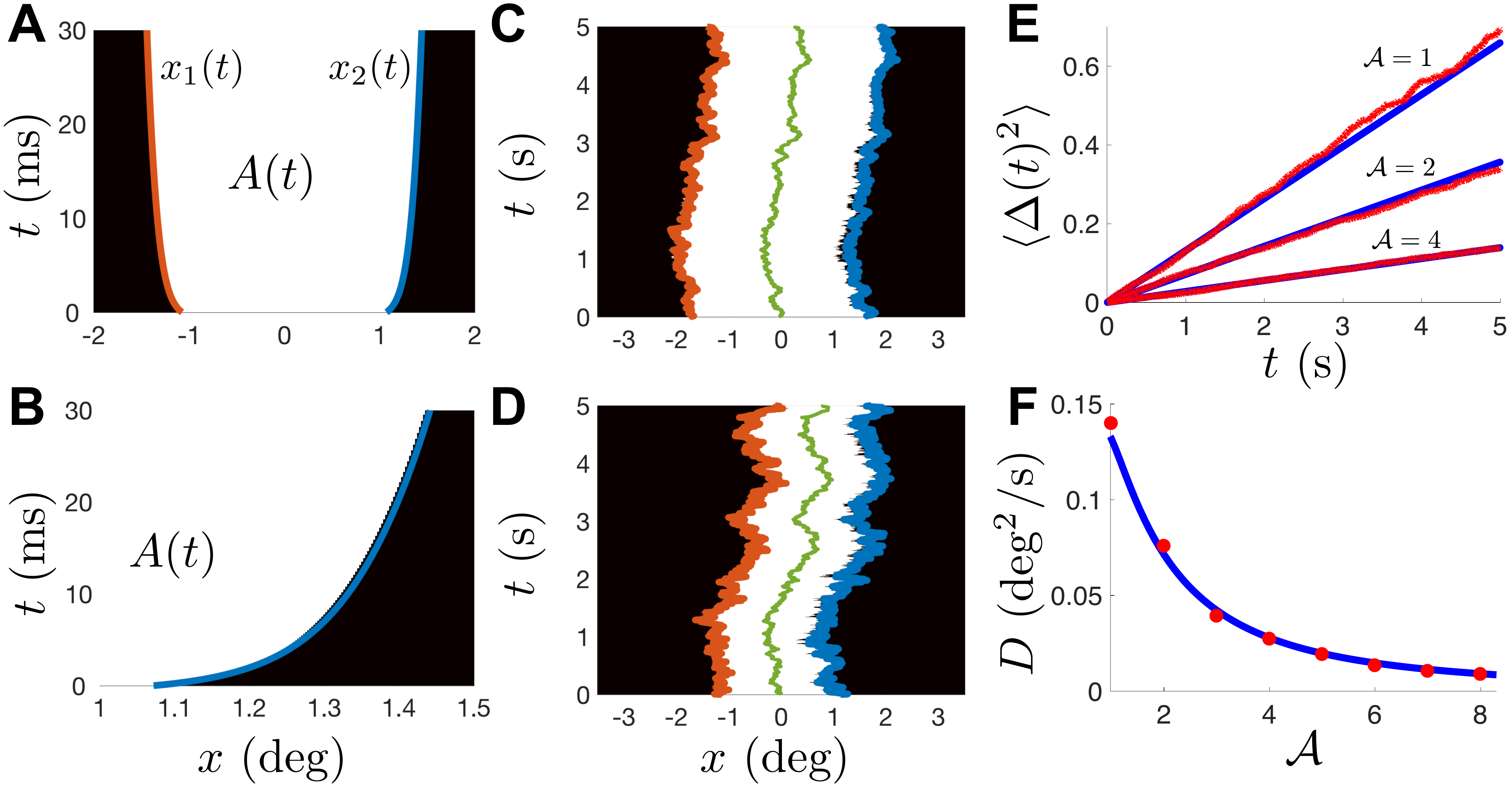} \end{center}
\caption{Low-dimensional interface equations, Eq.~(\ref{ifaceb}), approximate the dynamics of Eq.~(\ref{nfield}) well for single-bump solutions. {\bf A}. Bump of the form $u(x,0) = 0.25 \cdot U_0(x)$, Eq.~(\ref{bumpsol}), expands outward to the equilibrium shape $u(x,t) \to U_0(x)$ for ${\mc A}=2$. Interface dynamics $u(x_{1,2}(t),t) = \theta$ are well approximated by Eq.~(\ref{oneone}), computed for the left (red curve) and right (blue curve) interfaces. {\bf B}. Zoom-in of {\bf A}. {\bf C}. Bump initially of form $u(x,0) = U_0(x)$, for ${\mc A}=2$, evolves stochastically in response to noise with correlation structure Eq.~(\ref{cosnoise}) with $\ep = 0.03$. Interface dynamics approximated by Eq.~(\ref{stochint}), plotted for the left (red trajectory) and right (blue trajectory). Centroid (green trajectory) can be approximated by Eq.~(\ref{bcenteq}). {\bf D}. For ${\mc A}=1$, the bump is narrower and diffuses more in response to the noise. {\bf E}. Variance of the bump as a function of time. Theory (blue line) given by Eq.~(\ref{diffcon}) is compared with the results of $10^4$ numerical simulations (red line) of Eq.~(\ref{nfield}). {\bf F}. Diffusion coefficient computed as a function of $\A$. Theory (solid line) and sims (dots) compare favorably, showing a systematic decrease in diffusion as the strength of coupling $\A$ is increased. Threshold $\theta = 0.25$ throughout, and $\omega_c = 25 \pi / 180$. Numerical simulations are performed using Euler-Maruyama stochastic integration scheme with $dx = 0.005$ and $dt = 0.1$.}
\label{fig4_onebump}
\end{figure*}
Lastly, note that we can examine the effects of noise by analyzing the stochastic differential equations
\begin{subequations} \label{stochint}
\begin{align}
\d x_1(t) &= \frac{\theta - W(x_2(t) - x_1(t))}{\bar{\alpha}} \d t \\
& \hspace{18mm} - \frac{\sqrt{\ep \cdot \theta}}{\bar{\alpha}} \d Z(x_1(t),t), \nonumber  \\
\d x_2(t) &= \frac{- \theta + W(x_2(t) - x_1(t))}{\bar{\alpha}} \d t \\
& \hspace{18mm} + \frac{\sqrt{\ep \cdot \theta} }{\bar{\alpha}} \d Z(x_2(t),t).   \nonumber
\end{align}
\end{subequations}
As in previous work~\citep{kilpatrick13,carroll14}, we can track the stochastic motion of the bump's location by looking at the center of mass $\Delta (t) = (x_1(t) + x_2(t))/2$, evolving as
\begin{align}
\d \Delta (t) = \frac{\sqrt{\ep \cdot \theta}}{2 \bar{\alpha}} \left[ \d Z(x_2(t),t) - \d Z(x_1(t),t) \right]. \label{bcenteq}
\end{align} 
Assuming fluctuations in the active region width are small ($x_2-x_1 \approx 2h$), we approximate $x_1 \approx \Delta - h$ and $x_2 \approx \Delta + h$, so we can directly compute the mean $\langle \Delta (t) \rangle = \Delta (0)$ and variance $ \langle (\Delta(t) - \langle \Delta(t) \rangle )^2 \rangle = Dt$ with diffusion coefficient
\begin{align}
D &= \frac{\ep \theta}{4 \bar{\alpha}^2} \left[ \langle Z(\Delta - h,t)^2 \rangle + \langle Z(\Delta +h,t)^2 \rangle \right. \nonumber \\
& \hspace{17mm} \left. - 2 \langle Z(\Delta - h,t) Z(\Delta + h,t) \rangle \right], \nonumber \\
&= \frac{\ep \theta}{2 \bar{\alpha}^2} \left[ C(0) - C(2h) \right],  \label{diffcon}
\end{align}
where $C(x)$ is the spatial correlation function of the neural field, Eq.~(\ref{nfield}). We derived a related equation in the case of additive noise by directly assuming stochastic motion of the bump's position in \cite{kilpatrick13}. Note, Eq.~(\ref{diffcon}) provides a formula for the diffusion coefficient of the bump for arbitrary spatial noise correlations $C(x)$, in contrast to the work of \cite{burak12}, which assumes independent fluctuations are generated at each point in the network via a Poisson process.

We demonstrate the accuracy of these approximations by tracking the transient evolution of bumps in numerical simulations and comparing with predictions of Eq.~(\ref{ifaceb}). First, for a bump unforced by noise that is initiated with a narrower width than its equilibrium width, given by the wide solution to Eq.~(\ref{widform}), the interfaces relax outward. In fact, these dynamics can be tracked by the interface Eq.~(\ref{oneone}), corresponding to the half-width of the evolving bump. We need only calculate the antiderivative Eq.~(\ref{wanti}) for our specific weight function, Eq.~(\ref{wizard}), given by Eq.~(\ref{wizanti}). Thus, we can compute from Eq.~(\ref{bumpsol}) that
\begin{align}
\bar{\alpha} = |U_{0}'(\pm h)| = \A \left[ 1 - (1- 2h (\A) )\e^{-2h (\A)} \right], \label{alform}
\end{align}
where the half-width depends on the spatial scale $h(\A)$. In Fig. \ref{fig4_onebump}A,B, we compare the level sets $u(x_j(t),t) = \theta$ of a numerical simulation of Eq.~(\ref{nfield}) initiated with a narrower initial condition to the evolution of the interface Eq.~(\ref{oneone}), showing the expansion is tracked well. We also use Eq.~(\ref{alform}) to approximate the evolution of noise-driven bumps, described by the stochastic interface Eq.~(\ref{stochint}). We use a cosine spatial correlation function
\begin{align}
C(x) = \cos (\omega_c x), \ \ \ \ \omega_c = c \pi /L. \label{cosnoise}
\end{align}
We demonstrate agreement between the prediction of our interface Eq.~(\ref{stochint}) and the level sets of Eq.~(\ref{nfield}) of a bump perturbed by noise in Fig. \ref{fig4_onebump}C,D. Note, for consistency the noise increments $\d Z(x,t)$ generated for the neural field model, Eq.~(\ref{nfield}), are used to generate the noise perturbations $\d Z(x_j,t)$ for the interface Eq.~(\ref{stochint}). A similar approach is used in subsequent single realizations shown in Figs. \ref{fig7_twostoch} and \ref{fig10_multisim}. Furthermore, we can approximate the diffusion coefficient $D$ corresponding to the rate at which the variance of the bump's centroid grows $ \langle (\Delta(t) - \langle \Delta(t) \rangle )^2 \rangle = Dt$, as given by Eq.~(\ref{diffcon}), so
\begin{align}
D = \frac{\ep \theta}{2 {\mc A}^2} \frac{\D 1 - \cos (2 \omega_c h) }{\left( 1 + (2h ({\mc A}) - 1) \e^{-2h ({\mc A})} \right)^2}.  \label{difcomp}
\end{align}
similar to results derived in \citet{bressloff12b,kilpatrick13}.
%Note that as $\gamma$ is decreased, the spatial scale of the noise increases, so correlations spread farther across the domain. Since the noise perturbations to each bump interface become more correlated, the diffusion coefficient decreases as a result. We can show this by differentiating $D$ with respect to $\gamma$ to yield
%\begin{align*}
%\frac{\d D}{\d \gamma} = \frac{\ep}{2 \A^2 } \frac{\D 4 \gamma h({\mc A})^2}{\left( 1 - (1-2h (\A) ) \e^{-2h (\A)} \right)^2} \e^{-2 \gamma h(\A)} > 0, 
%\end{align*}
%for $\gamma >0$, so $D$ decreases as $\gamma$ is decreased.
Fixing the noise amplitude $\ep$, we find the diffusion coefficient decreases monotonically as the synaptic strength $\A$ is increased (Fig. \ref{fig4_onebump}E,F). Our analytical approximation, Eq.~(\ref{difcomp}), agrees well with simulations of the full model, Eq.~(\ref{nfield}). Thus, stronger synaptic inputs (larger ${\mc A}$) increase the size of the bumps (larger $h$), and these wider bumps are more stable to noise perturbations (smaller $D$).

% Section 3
\section{Dynamics of two interacting bumps}
\label{twobumps}

Our main interest lies in understanding how multiple bumps interact, as these interactions will contribute to our model's limitations in multi-item WM. Prior to examining WM for an arbitrary number of items, we focus on the case of two interacting bumps, to demonstrate how bump interactions lead to different errors during the delay period of a WM trial. The effective equations for two bumps can then be extended to higher dimensions.

\subsection{Interface equations}
We begin by extending our interface Eq.~(\ref{ifaceb}) for one bump to the case of two bumps.  Again, since the nonlinearity in Eq.~(\ref{nfield}) is a step function $H(u-\theta)$, its output is determined by the active region $A(t) = \{ x | u(x,t) \geq \theta \}$.
%This allows one to rewrite Eq.~(\ref{nfield}) (for $Z \equiv 0$) as
%\begin{align}
%\d u(x,t) &= \left[ - u(x,t) + \int_{A(t)} w(x-y) \d y + I(x,t). \label{afield}
%\end{align}
In the case of a single bump, we defined a simply connected region as in Eq.~(\ref{nfact}). Two bumps would typically be comprised of two disjoint active regions. However, if the bumps began close enough together, their active regions would overlap and form a single connected domain. In this special case, the dynamics of the system would subsequently be described by the single bump interface Eq.~(\ref{ifaceb}). Thus, for our analysis here, we assume the active region is comprised of two disjoint subdomains, $A(t) = \left[ x_1(t),x_2(t) \right] \cup \left[ x_3(t), x_4(t) \right]$, so Eq.~(\ref{nfield}) becomes
\begin{align}
&\d u(x,t)= \left[ -u(x,t) + \int_{x_1(t)}^{x_2(t)} w(x-y) \d y \right. \label{twofield} \\
&  \left. + \int_{x_3(t)}^{x_4(t)} w(x-y) \d y \right] \d t + \sqrt{\ep \cdot |u(x,t)|} \d Z(x,t).   \nonumber
\end{align}
As mentioned, we assume two bumps have been instantiated far enough apart so that their active regions do not overlap. Assuming continuity of $u(x,t)$, the boundary points of $A(t)$ correspond to the interfaces of the bumps, and satisfy the dynamic threshold equations
\begin{align}
u(x_j(t),t) = \theta, \hspace{4mm} j=1,2,3,4.  \label{dthresh}
\end{align}
Differentiating Eq.~(\ref{dthresh}) with respect to $t$, we find the total derivative is again given by Eq.~(\ref{ifcon}). Specifying the integral terms in Eq.~(\ref{twofield}) using Eq.~(\ref{wanti}),
%and noting that since $w(x)$ is even ($w(-x) = w(x)$), then $W(x)$ is odd ($W(-x) = -W(x)$),
we can rewrite integrals ${\mc I}$ as
\begin{align}
{\mc I}: &= \int_{x_1}^{x_2} w(x_j-y) \d y + \int_{x_3}^{x_4} w(x_j-y) \d y \nonumber \\
&=  \int_{x_j-x_2}^{x_j-x_1} w(z) \d z + \int_{x_j-x_4}^{x_j-x_3} w(x_j-y) \d y \nonumber \\
%&= W(x_j-x_1)-W(x_j-x_2) + W(x_j-x_3)-W(x_j-x_4) \nonumber \\
&= \sum_{k=1}^4 (-1)^{k-1} W(x_j(t)-x_k(t)).  \label{wintsum}
\end{align}

%We can obtain a differential equation for the dynamics of the interfaces $x_j(t)$:
%\begin{align}
%\frac{\d x_j}{\d t} = - \frac{1}{\alpha_j(t)} \left[ \int_{A(t)} w(x_j(t)-y) \d y - \theta \right], \hspace{4mm} j=1,2,3,4,  \label{iface1}
%\end{align}
%where we have substituted Eq.~(\ref{afield}) at $x_j(t)$ for $\frac{\pd u(x_j(t),t)}{\pd t}$.

%We can further specify Eq.~(\ref{iface1}) by defining the integral
%\begin{align}
%W(x) = \int_0^x w(y) \d y, \label{wint}
%\end{align}
%and it is useful to note that since $w(x)$ is even ($w(-x) = w(x)$), then $W(x)$ is odd ($W(-x) = -W(x)$). Subsequently, we can rewrite the integral terms in all four Eq.~(\ref{iface1}) as
\begin{figure*}
\begin{center} \includegraphics[width=13cm]{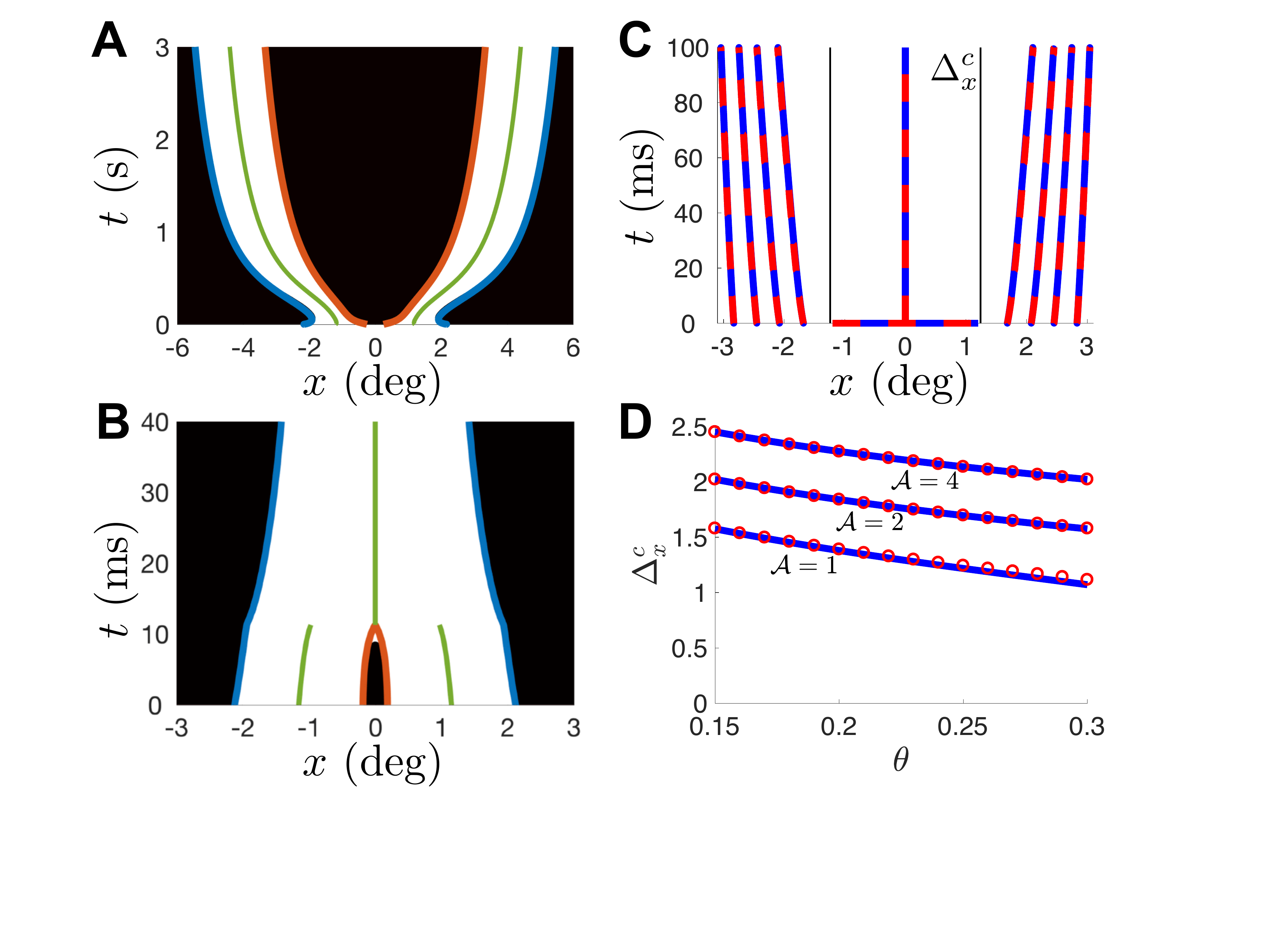} \end{center}
\caption{Two bumps interacting in the deterministic ($Z \equiv 0$) neural field Eq.~(\ref{nfield}). {\bf A}. Two bumps repel each other when initiated at $\pm x_0 = \pm 1.25$. The location of the bump interfaces are well-tracked by the curves (solid lines) generated by the low-dimensional Eq.~(\ref{twosym}). Parameters are $\theta = 0.25$ and ${\mc A}=1$. {\bf B}. Two bumps merge when initiated at $\pm x_0 = \pm 1.23$, a short distance for the point of initiation in {\bf A}. {\bf C}. Multiple trajectories of the centroid of the bump for initial conditions $x_0 = \{1.2, 1.6, 2.0, 2.4, 2.8\}$ as predicted by Eq.~(\ref{twosym}) (solid line) and direct simulation (dashed lines). Bumps repel each other more strongly when they begin close to the critical boundary $\Delta_x^c$ (thin line). Within the critical boundary, bumps merge. {\bf D}. The critical boundary $\Delta_x^c(\theta)$ is determined by Eq.~(\ref{bmerge}) (solid curves), and compared with direct simulations (circles). See Fig. \ref{fig4_onebump} for details on numerical simulations.}
\label{fig5_twoiface}
\end{figure*}

We study two cases of the two bump interface equations, which admit different approximations. For a fully deterministic Eq.~(\ref{nfield}), we can derive an integral equation for the time-evolution of the spatial gradients $\alpha_j(t)$ at the interfaces. An alternative approach, which is more straightforward, is to simply approximate the gradients $\alpha_j(t) \equiv \bar{\alpha}_j$ using static quantities derived from stationary solutions of Eq.~(\ref{nfield}). This is easier to employ, especially in the case of stochastic forcing. \\
\vspace{-2mm}

\noindent
{\bf Dynamic gradients.} In the case of a fully deterministic system ($Z \equiv 0$ in Eq.~(\ref{nfield})), we can follow \citet{coombes12} to obtain an analytic formula for $\alpha_j(t)$ by defining $z(x,t) : = \frac{\pd u(x,t)}{\pd x}$ and differentiating Eq.~(\ref{twofield}) with respect to $x$ to find
\begin{align*}
\frac{\pd z(x,t)}{\pd t} &= - z(x,t) + \sum_{k=1}^4 (-1)^{k-1} w(x-x_k(t)),
\end{align*}
which we can integrate to yield
\begin{align}
z(x,t) &= \e^{-t} \int_0^{t} \e^s \left[ \sum_{k=1}^4 (-1)^{k-1} w(x-x_k(s))  \right]  \d s \nonumber \\
& \hspace{10mm} + z(x,0) \e^{-t}. \label{zeqn}
\end{align}
Evaluating Eq.~(\ref{zeqn}) at $x_j(t)$, we have:
\begin{align}
\alpha_j(t) &=  \int_{0}^{t} \e^{-(t-s)} \left[ \sum_{k=1}^4 (-1)^{k-1} w(x_j(t)-x_k(s))  \right]  \d s \nonumber \\
& \hspace{10mm} + u_0'(x_j(t))\e^{-t}, \ \ j=1,2,3,4, \label{alpha1}
\end{align}
where $u(x,0) = u_0(x)$ is the initial condition. Thus, we have a closed system describing the evolution of the interfaces of the two stationary bumps, assuming the active region $A(t)$ remains as two disjoint subdomains
\begin{align}
\frac{\d x_j}{\d t} = - \frac{1}{\alpha_j(t)} \left[  \sum_{k=1}^4 (-1)^{k-1} W(x_j(t)-x_k(t)) - \theta  \right]   \label{iface1}
\end{align}
for $ j=1,2,3,4$, with $\alpha_j(t)$ defined as in Eq.~(\ref{alpha1}). The second order term in Eq.~(\ref{ifcon}) will vanish, since there is no noise in this case. As we have performed no truncations, the pair of Eq.~(\ref{alpha1}) and (\ref{iface1}) exactly characterize the motion of the four bump interfaces $(x_1,x_2,x_3,x_4)$. We compare the evolution of the interfaces given by Eqs.~(\ref{alpha1}) and (\ref{iface1}) to those calculated from the full model Eq.~(\ref{nfield}) in Fig. \ref{fig5_twoiface}A,B. Bumps can either move away from each other (Fig. \ref{fig5_twoiface}A) or towards each other (Fig. \ref{fig5_twoiface}B), depending on the initial distance $2 \Delta_x : = (x_4(0)+x_3(0) - x_2(0) - x_1(0))/2$ of their centroids from one another. Note, merging occurs extremely rapidly, and repulsion happens otherwise. We can characterize the relaxation rate of merging by linearizing Eq.~(\ref{oneone}) about the equilibrium bump half-width $h$. As shown in previous stability analyses of stationary bumps in Eq.~(\ref{nfield}), the eigenvalue associated with the decay of width perturbations is $\lambda_e = 2w(2h)/(w(0) - w(2h))< 0$~\citep{amari77,kilpatrick13}. In our performance calculations, we account for this, and ignore the detailed dynamics of merging.

Note that for initial conditions that are symmetric about $x=0$ ($x_4(0) = -x_1(0) = b(0)$ and $x_3(0) = -x_2(0) = a(0)$), the interfaces evolve symmetrically: $x_4(t) = -x_1(t) = b(t) \geq 0$ and $x_3(t) = -x_2(t) = a(t) \geq 0$. In a similar way, the dynamic gradients exhibit odd symmetry: $\alpha_1(t) = -\alpha_4(t) = \beta (t) \geq 0$ and $\alpha_3(t) = - \alpha_2(t) = \alpha (t) \geq 0$. Eqs.~(\ref{alpha1}) and (\ref{iface1}) can be reduced to four equations:
\begin{subequations} \label{twosym}
\begin{align}
&\frac{\d a(t)}{\d t} = \frac{1}{\alpha (t)} \left[ \theta - W(b (t) - a (t)) + W(2a(t)) \right. \nonumber \\
& \hspace{15mm} \left. - W(a(t) + b(t)) \right],  \label{twosyma} \\
& \frac{\d b(t)}{\d t}  = \frac{1}{\beta (t)} \left[ W(b(t) - a(t)) - \theta + W(2b(t)) \right. \nonumber \\
& \hspace{15mm} \left. - W(a(t) + b(t)) \right], \\
& \alpha (t)  = \e^{-t} \int_0^t \e^s \left[ w(a(t)+b(s)) - w(a(t)+a(s)) \right. \nonumber \\
& \hspace{0mm} \left. + w(a(t)-a(s)) - w(a(t)-b(s)) \right] \d s + u_0'(a(t)) \e^{-t}, \\
& \beta (t)  = \e^{-t} \int_0^t \e^s \left[ w(b(t)+a(s)) -w(b(t) + b(s))   \right. \nonumber \\
& \hspace{0mm} \left. + w(b(t) - b(s)) - w(b(t)-a(s))  \right] \d s - u_0'(b(t)) \e^{-t}.
\end{align}
\end{subequations}
The system, Eq.~(\ref{twosym}), is used to calculate the interfaces of the two scenarios shown in Fig. \ref{fig5_twoiface}A,B.

\begin{figure}
\begin{center} \includegraphics[width=6cm]{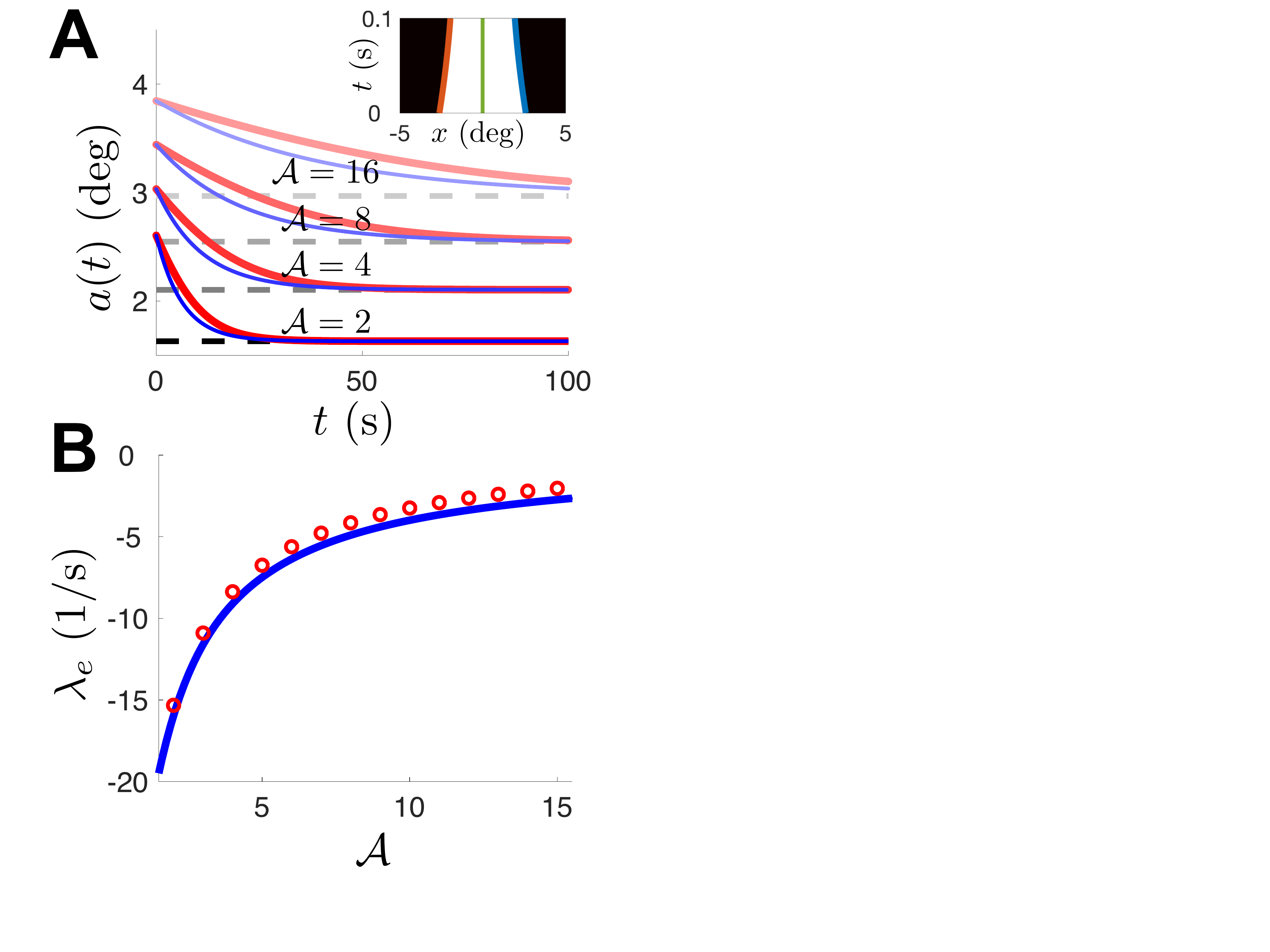} \end{center}
\vspace{-2mm}
\caption{Decay rate of merged bumps to equilibrium bump shape. {\bf A}. Two bumps are initiated at $\pm x_0  = \pm 1.23$, leading to an initial overlap in their active regions (Inset shows simulation of Eq.~(\ref{nfield}) for ${\mc A} =2$). The resulting merged bump has initial half-width $a(0)$ which decays towards the equilibrium half-width $h({\mc A})$ (dashed lines). Interface dynamics $u(\pm a(t),t) = \theta$ are computed directly (red lines) from Eq.~(\ref{nfield}) and by approximation $a(t) \approx a(0) + (a(0)-h)\e^{\lambda_e t}$ (blue lines) where $\lambda_e$ is given by Eq.~(\ref{eveig}). {\bf B}. The rate of decay ($|\lambda_e|$) decreases as ${\mc A}$ is increased, meaning networks with stronger synapses have bumps whose widths decay more slowly. Analytical approximation (blue line) using Eq.~(\ref{eveig}) compares well with decay rate of best exponential fit to $a(t)$ in {\bf A} (red circles). Threshold $\theta = 0.25$. See Fig. \ref{fig4_onebump} for details of numerical simulations.}
\label{fig6_decay}
\vspace{-2mm}
\end{figure}

There is a critical distance $2\Delta_x^c$ between two bumps initial centroids, which divides solutions that repel ($\Delta_x> \Delta_x^c$) from those that merge ($\Delta_x < \Delta_x^c$). We illustrate this by tracking the centroids of two symmetrically placed bumps for various starting distances $\Delta_x$ (Fig. \ref{fig5_twoiface}C). Similar features of associated spiking network models have been identified in \citet{wei12,almeida15}. We can determine an analytical expression that accurately characterizes the critical distance $\Delta_x^c$. Utilizing Eq.~(\ref{twosyma}), we note that if $a'(t) < 0$, bumps will initially move towards one another. Motivated by the findings of our numerical simulations in Fig. \ref{fig5_twoiface}C, we expect bumps that are initially attracted will continue to move towards one another until they merge. In this case, the critical curve $(a^c,b^c)$, determined by the condition
\begin{align}
\theta = W(b^c - a^c) - W(2a^c) + W(a^c + b^c) \label{bmerge}
\end{align}
divides initial conditions $(a(0),b(0))$ that merge from those that repel each other. Assuming the bumps initially have width $b(0) - a(0) = 2h$, as prescribed by Eq.~(\ref{bumpsol}), then $\theta = W(b(0) - a(0))$ and Eq.~(\ref{bmerge}) simplifies to $W(2 a^c) = W(a^c+b^c)$. Thus, defining the right bump's initial centroid $\Delta_x^c = (a^c+b^c)/2$, then $a^c = \Delta_x^c -h$ and $b^c = \Delta_x^c + h$, so $W(2 \Delta_x^c - 2h) = W(2 \Delta_x^c)$. Applying Eq.~(\ref{wanti}) and simplifying, we find $\Delta_x^c = h/(1 - \e^{-2h})$, which is increasing in $h$ for $h>1/2$. Thus, as expected, wider bumps will always have a wider critical merging distance $\Delta_x^c$. We compare our analytical prediction, Eq.~(\ref{bmerge}), to the results of numerical simulations, and find they agree (Fig. \ref{fig5_twoiface}D). Thus, our interface equations not only track the motion of bumps, but can also predict when they merge with one another.

For the special case in which symmetric bumps initially overlap, merging occurs immediately, and the subsequent interface dynamics are well approximated by Eq.~(\ref{oneone}). In this case, the associated dynamics is given by the decay of the width of the resulting merged bump (inset in Fig. \ref{fig6_decay}A). We can linearly approximate the decay dynamics of the interfaces, where $u(\pm a(t),t) = \theta$, as $a(t) \approx h + (a(0)-h) \e^{\lambda_et}$, with $\lambda_e$ the eigenvalue in Eq.~(\ref{eveig}). This approximation agrees well with simulations of the full model Eq.~(\ref{nfield}) for a wide range of synaptic strengths ${\mc A}$ (Fig. \ref{fig6_decay}A). Interestingly, the rate ($|\lambda_e|$) at which the bump width decays decreases as ${\mc A}$ is increased (Fig. \ref{fig6_decay}B). Thus, it takes longer for bumps in networks with strong synapses to reach their equilibrium half-width $h$. \\
\vspace{-3mm}

\begin{figure*}
\begin{center} \includegraphics[width=11cm]{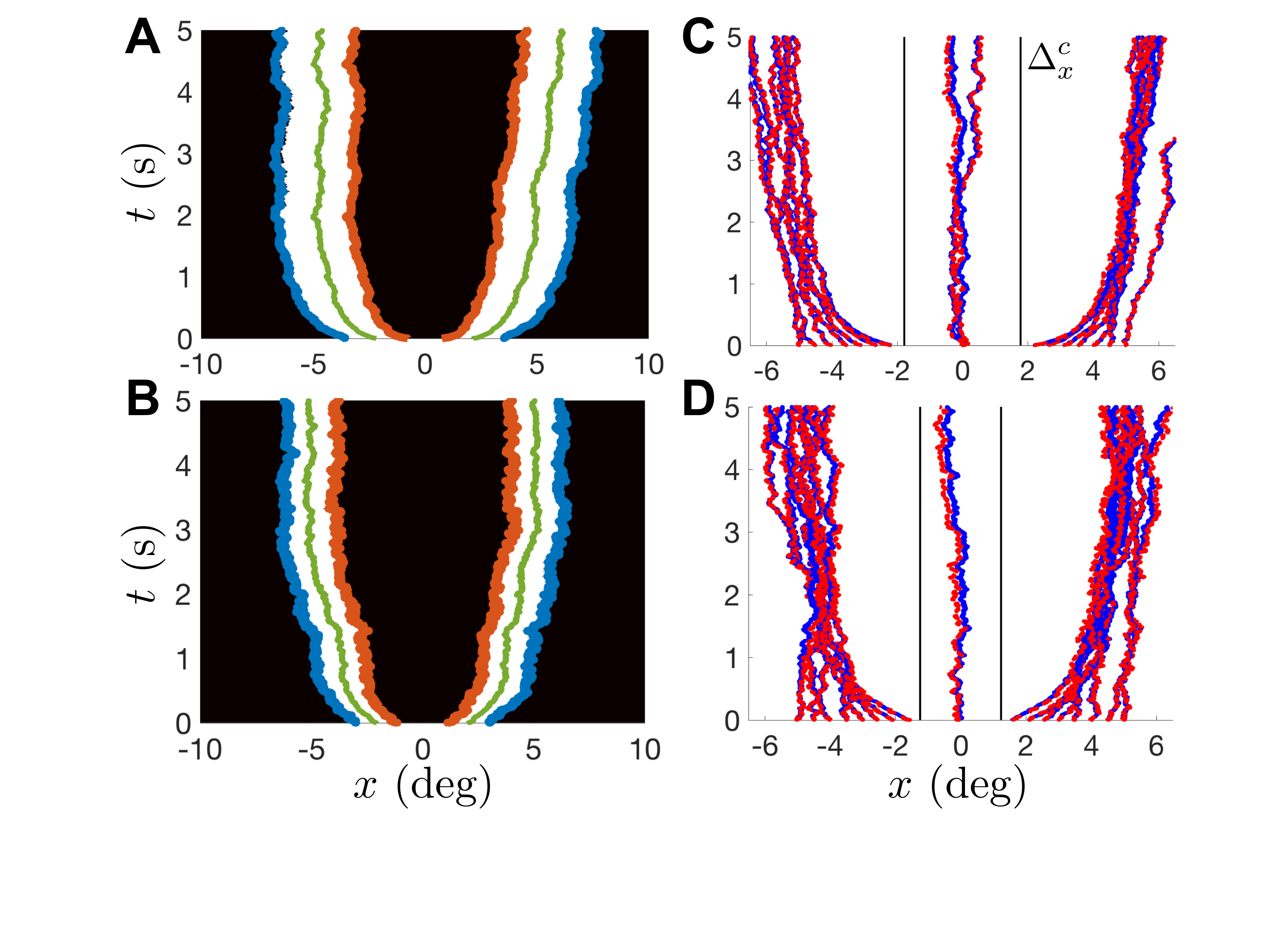} \end{center}
\caption{Stochastic simulations of Eq.~(\ref{nfield}) with noise amplitude $\ep = 0.03$ and noise correlations, Eq.~(\ref{cosnoise}). {\bf A}. Two noise-driven bumps repel each other and diffuse when starting at $\pm x_0 = \pm 2$. Our estimate using the static gradient approximation (solid lines), Eq.~(\ref{twostat}), tracks the interfaces. Other parameters are $\theta = 0.25$ and ${\mc A}=2$. {\bf B}. When ${\mc A}=1$, the bumps repel each other less than in {\bf A}. {\bf C}. Trajectories of the centroid of the noise-driven bumps with centroids initiated at different locations $\pm x_0$ from 1 to 5 deg, spaced 0.5 deg apart. Approximations (solid lines) using the low-dimensional system, Eq.~(\ref{twocent}), agree with direct numerical simulations (dashed lines). When initiated within the region $x_0 < \Delta_x^c$ (thin lines), bumps merge as in the deterministic Fig. \ref{fig5_twoiface}. Here, ${\mc A}=2$. {\bf D} Same as {\bf C}, except ${\mc A}=1$, showing bumps repel each other less and also merge in a narrower range. See Fig. \ref{fig4_onebump} for details on numerical simulations.}
\label{fig7_twostoch}
\end{figure*}

\noindent
{\bf Static gradient approximation.} An alternative to computing the integral equations in Eq.~(\ref{alpha1}) is to assume a static approximation to the gradients $\alpha_j(t) \equiv \bar{\alpha}_j$. Since our solutions evolve with a profile approximated by sums of the stationary bump solution $U_{0}(x)$, Eq.~(\ref{bumpint}), we consider gradients approximated by $U_{0}'(x)$. For the weight function Eq.~(\ref{wizard}), these can be computed directly from Eq.~(\ref{alform}) as
\begin{align*}
\bar{\alpha}_{1,3} = -\bar{\alpha}_{2,4} = \bar{\alpha} = U_{0}'(-h).
\end{align*}
We can then write the resulting interface equations in a simple form that still captures the interactions between the two bumps, as well as the effects of noise perturbations:
\begin{subequations} \label{twostat}
\begin{align}
\d x_1 &= \frac{1}{\bar{\alpha}} \left( \left[ \theta- W(x_2 - x_1) + W(x_3 - x_1) \right. \right. \\
& \hspace{3mm} \left.  \left. - W(x_4 - x_1) \right] \d t - \sqrt{\ep \cdot \theta} \hspace{0.5mm} \d Z(x_1,t) \right), \nonumber \\
\d x_2 &= -\frac{1}{\bar{\alpha}} \left( \left[ \theta- W(x_2 - x_1) + W(x_3 - x_2) \hspace{-2mm} \right. \right. \\
& \hspace{3mm} \left. \left.  - W(x_4 - x_2) \right] \d t - \sqrt{\ep \cdot \theta} \hspace{0.5mm} \d Z(x_2,t) \right), \nonumber  \\
\d x_3 &= \frac{1}{\bar{\alpha}} \left( \left[ \theta- W(x_4 - x_3) + W(x_3 - x_2) \hspace{-2mm} \right. \right. \\
& \hspace{3mm} \left. \left.  - W(x_3 - x_1) \right] \d t - \sqrt{\ep \cdot \theta} \hspace{0.5mm} \d Z(x_3,t) \right), \nonumber \\
\d x_4 &= -\frac{1}{\bar{\alpha}} \left( \left[ \theta- W(x_4 - x_3) + W(x_4 - x_2) \hspace{-2mm} \right. \right. \\
& \hspace{3mm} \left. \left. - W(x_4 - x_1) \right] \d t - \sqrt{\ep \cdot \theta} \hspace{0.5mm} \d Z(x_4,t) \right). \nonumber
\end{align}
\end{subequations}
As in the case of single-bumps, the second order term in Eq.~(\ref{ifcon}) is smaller than ${\mc O}(\sqrt{\ep})$. The system Eq.~(\ref{twostat}) accurately approximatess the stochastic dynamics of the two bumps' interfaces (Fig. \ref{fig7_twostoch}A,B). We can also reduce Eq.~(\ref{twostat}) to track the centroid of each bump $\Delta_1 = (x_1 + x_2)/2$ and $\Delta_2 = (x_3 + x_4)/2$. To do so, we assume the width of each bump remains approximately constant, so $x_1 \approx \Delta_1 - h$, $x_2 \approx \Delta_1 +h$, $x_3 \approx \Delta_2-h$, and $x_4 \approx \Delta_2 + h$. In this case, we find two equations for the stochastic dynamics of the centroids:
\begin{subequations} \label{twocent}
\begin{align}
\d \Delta_1 & = \frac{1}{\bar{\alpha}} \left( J(\Delta_2 - \Delta_1) \d t + \frac{\sqrt{\ep \cdot \theta}}{2} \hspace{0.5mm} \d {\mc Z}(\Delta_1,t) \right), \\
\d \Delta_2 &= \frac{1}{\bar{\alpha}} \left( J(\Delta_1 - \Delta_2) \d t + \frac{\sqrt{\ep \cdot \theta}}{2} \hspace{0.5mm} \d {\mc Z}(\Delta_2,t) \right),
\end{align}
\end{subequations}
where odd symmetry of the coupling function $J(\Delta)$ follows from the odd symmetry of $W(x) = -W(-x)$:
\begin{align*}
& J(\Delta) = \frac{1}{2} \left( 2W(\Delta) - W(\Delta - 2h) - W(\Delta + 2h) \right), \\
& \d {\mc Z}(\Delta_j,t) = \d Z(\Delta_j+h,t) - \d Z(\Delta_j - h,t),
\end{align*}
for $j=1,2$. We use the approximation, Eq.~(\ref{twocent}), to determine the evolution of the centroids in realizations with different initial conditions (Fig. \ref{fig7_twostoch}C,D). Even though the bump for ${\mc A}=2$ has a lower diffusion coefficient (as shown in Fig. \ref{fig4_onebump}E,F), for ${\mc A}=1$, bumps repel each other less. Thus, bumps in networks with weaker synaptic weights can stray less from their initial position, when they are initiated close together. However, when bumps begin far apart, strong connectivity may be more advantageous, since bumps are less perturbed by noise. We examine this tradeoff by determining performance of the network in a two-item WM task, using our approximations and full numerical simulations.

\subsection{Performance}

We study estimation errors of the network encoding locations of two targets, $\phi_1$ and $\phi_2$. In general, we categorize errors as arising from (a) merging; (b) repulsion; and (c) diffusion of the bumps encoding these targets. The fluctuation-driven random walk of bumps has been characterized in single-item WM models~\citep{compte00,kilpatrick13}, and validated in behavioral and electrophysiological experiments~\citep{wimmer14,constantinidis16}. Bump merging was recently characterized in spiking network models of multi-item WM~\citep{wei12,almeida15}, motivated by corresponding human psychophysics data~\citep{bays09}. The merging of item memories has been considered in heuristic models of multi-item WM, and analyses sometimes assume the memory for one of the associated items is then completely lost, so subjects guess to report its location~\citep{zhang08}. Similar guesses may occur due to attentional lapses, where subjects do not store an item in the first place. We avoid such characterizations in our analysis, and study error solely ascribed to the dynamics of bump attractors in Eq.~(\ref{nfield}). Thus, we assume that when bumps merge, the remaining bump encodes the location of both items corresponding to the original two bumps. Finally, note that repulsion will lead to item memories that diverge from one another when bumps are instantiated close to one another.

\begin{figure*}
\begin{center} \includegraphics[width=15.2cm]{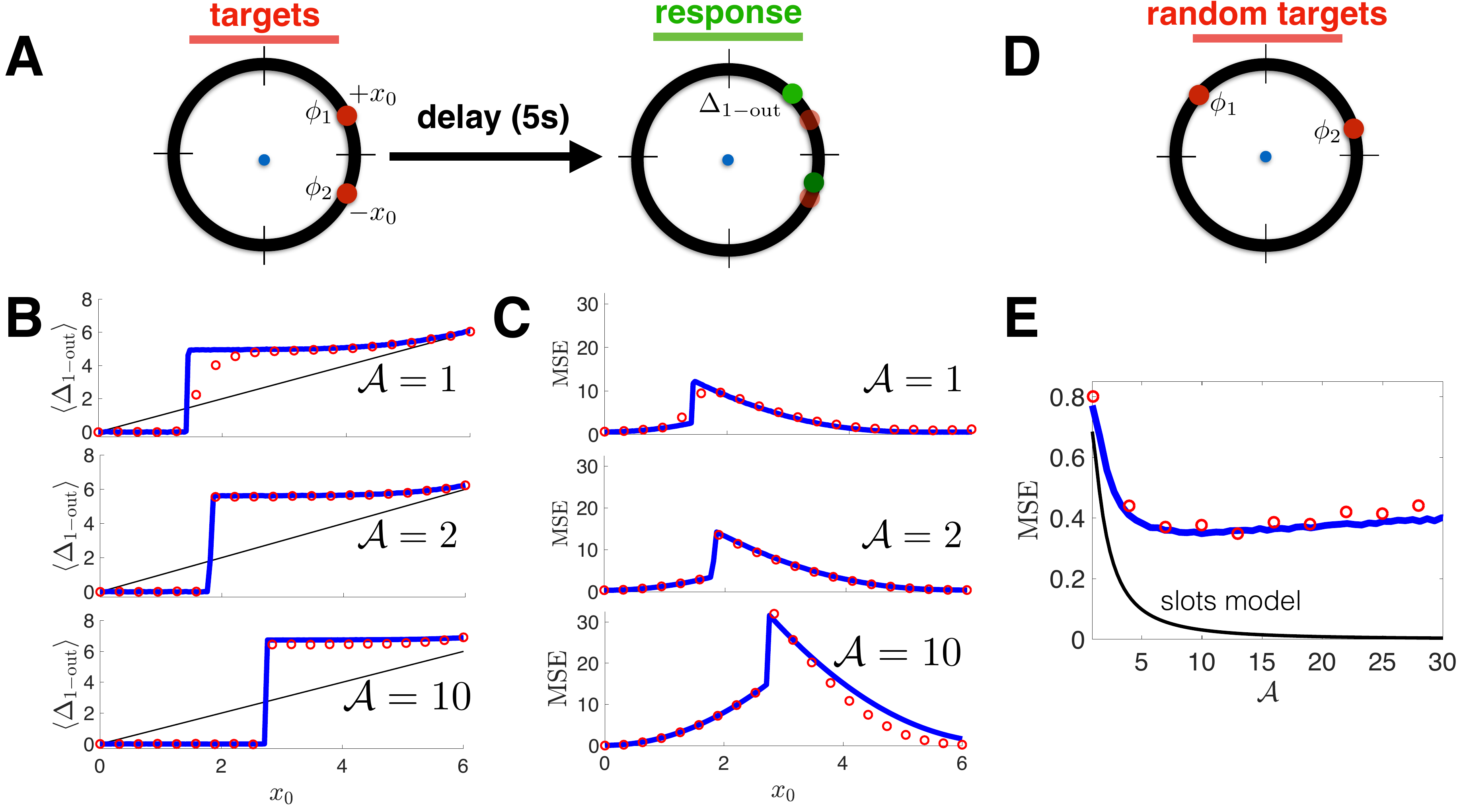} \end{center}
\vspace{-2mm}
\caption{Recall error in a two-item WM task. {\bf A}. Item-distance dependence of errors. Target items $\phi_{1,2}$ are located at $\pm x_0$. Mean squared error (MSE) is computed by comparing output angle $\Delta_{1-\text{out}}$ to input $\phi_{1}$ via Eq.~(\ref{msedef}). {\bf B}. Mean location of $\Delta_{1-\text{out}}$ computed as a function of $x_0$ for low-dimensional Eq.~(\ref{twocent}), (solid lines), and full simulations of the neural field (circles). Merging leads to values close to zero. A transition occurs at the boundary of merging, where repulsion begins. For large enough $x_0$, $\langle \Delta_{1-\text{out}} \rangle \approx +x_0$. Compare ${\mc A}=1, 2, 10$. {\bf C}. MSE as a function of $x_0$ computed from Eq.~(\ref{msedef}) is maximized near the boundary point, where repulsion is strongest, decreasing as the bumps are placed further apart. {\bf D}. MSE is also computed for uniform random targets $\phi_{1,2}$, averaging across all locations to compute the MSE for each synaptic strength ${\mc A}$.  {\bf E}. MSE is nonmonotonic in ${\mc A}$, with fluctuation-driven errors dominating for low ${\mc A}$ and merging/repulsion errors dominating for large ${\mc A}$. MSE is much larger would be predicted for a slots model, which assumes MSE is unchanged between one and two item memory tasks, so we use ${\rm MSE}_{\rm slots} = D \cdot T$ where $D$ is given by Eq.~(\ref{difcomp}). Blue curves and red circles are generated from $10^6$ Monte Carlo simulations.}
\label{fig8_twoperf}
\vspace{-4mm}
\end{figure*}

Across multiple trials, the task on trial $k$ is to encode both target angles $\phi_1^k,\phi_2^k \in [-180,180)$ ($\phi_1^k> \phi_2^k$). However, only a single target is probed, for instance by asking the subject to recall the angle corresponding to a particular color~\citep{wilken04,zhang08,bays09}. Due to the symmetry in the system, we compute the mean squared error (MSE) corresponding to the first target angle $\phi_1^k$ on each trial $k$
\begin{align}
\text{MSE} = \langle ( \Delta_{1\text{-out}} - \phi_1 )^2 \rangle = \frac{1}{K} \sum_{k=1}^{K} ( \Delta_{1\text{-out}}^k - \phi_1^k )^2, \label{msedef}
\end{align}
where $\Delta_{j\text{-out}}^1$ is the centroid of the bump encoding target $1$ at the end of the trial $k$ (as in Fig. \ref{fig8_twoperf}A,D).
%There is an important difference between our approach to defining error and that commonly used in psychophysics literature. Specifically, we compute the error as if a subject recalled both items in a single trial, but typical tasks only ask the subject to recall a single item~\citep{wilken04,bays09,zhang08}. Thus, our analysis essentially double counts each trial, when compared with psychophysics literature which will introduce small correlations. However, we expect the effect of this double counting will be small, and computing the error in our way provides more error measurements for the same number of trials, increasing the rate of our estimate's convergence.
The MSE in Eq.~(\ref{msedef}) can be computed directly from numerical simulations of Eq.~(\ref{nfield}), and we can also approximate the error using our reduced set of centroid equations, Eq.~(\ref{twocent}).

Our approximation of the MSE begins by determining whether or not the bumps merge. To do so, we examine the target distance to see if it is below the critical value, $(\phi_2 - \phi_1) < 2 \Delta_x^c$ (Fig. \ref{fig5_twoiface}D). As shown in Fig. \ref{fig5_twoiface}B, merging occurs very rapidly, so we do not model the detailed dynamics of merging in our performance calculations. Leveraging Eq.~(\ref{bmerge}), which describes the minimal distance at which symmetrically-placed bumps do not merge, we rotate coordinates of $\phi_{1,2}$ so they are symmetric about zero $\tilde{\phi}_{1,2} = (\phi_1-\phi_2, \phi_2-\phi_1)/2$. Thus, in Eq.~(\ref{bmerge}), assuming the initial bumps are roughly of width $2h$ we have $a^c = (\phi_2 - \phi_1-2h)/2$ and $b^c = (\phi_2 - \phi_1 + 2h)/2$, so if $W((\phi_2 - \phi_1-2h)/2)< W(\phi_2 - \phi_1)$, the bumps merge, otherwise they repel each other. This boundary is approximated given the weight function, Eq.~(\ref{wizard}), by solving the corresponding equality, $\phi_2 - \phi_1 = \frac{2h}{1 - \e^{-2h}}$, as the critical distance below which bumps merge. For this subset of cases, the MSE in Eq.~(\ref{msedef}) can be approximated in a straightforward way by noting
\begin{align}
{\mc M}_{\text{mg}} &= \langle ( \Delta_{1\text{-out}} - \phi_1 )^2 \rangle \nonumber \\
&= \langle ( \Delta_{1\text{-out}} - (\phi_1+\phi_2)/2 - (\phi_1-\phi_2)/2 )^2 \rangle \nonumber \\
&\approx  \langle ( \Delta_{1\text{-out}} - (\phi_1+\phi_2)/2)^2 \rangle + \langle (\phi_1-\phi_2)^2/4 \rangle  \nonumber \\
&= D \cdot T + \frac{(\phi_2-\phi_1)^2}{4},  \label{mergmse}
\end{align}
since the merged bump rapidly centers at the mean of the two target locations, $(\phi_1 + \phi_2)/2$. We can compute the diffusion coefficient $D$ using Eq.~(\ref{difcomp}), our theory for the stochastic dynamics of single bumps, and $T$ is the total delay time, providing us an analytic approximation of the MSE for the case $x_0 < \Delta_x^c$.

If bumps do not merge, we approximate their dynamics using the nonlinear stochastic system, Eq.~(\ref{twocent}).
The constituent function $J(\Delta)$ can be computed, given the weight function, Eq.~(\ref{wizard}),
\begin{align*}
J(\Delta) = -2 \A \e^{-\Delta} \left[ \Delta \sinh^2(h) - h \sinh(2h) \right],
\end{align*}
assuming $\Delta > 0$. Note, the formula for $J(\Delta)$ is more complicated for the case in which $\Delta_2< \Delta_1$, or their difference is across the periodic boundary at $x = \pm 180$. We consider these other cases in simulations, but do not discuss the formulas in detail here. In the case of noise correlations, Eq.~(\ref{cosnoise}), we can specify
\begin{align*}
\d {\mc Z}(\Delta_j,t) =& 2\sin(\omega_c h)  \left[  \sin(\omega_c \Delta_j) \cdot \d \xi_1(t)  \right. \\
&\hspace{15mm} \left. - \cos (\omega_c \Delta_j)  \cdot \d \xi_2(t) \right],
\end{align*}
where $\d \xi_j(t)$, $j=1,2$, are increments of a standard Weiner process. Eq.~(\ref{twocent}) is simulated numerically to estimate $\Delta_{1-\text{out}}^k$ in trial $k$, which then is plugged into our formula for MSE, Eq.~(\ref{msedef}).

Our approximations are compared to simulations of the full neural field model, Eq.~(\ref{nfield}), in Fig. \ref{fig8_twoperf}. Targets angles $\phi_1$ and $\phi_2$ are initially represented by instantiating a bump of form $U_0(x)$, Eq.~(\ref{bumpsol}), centered at the two locations in the neural field. Distance-dependence is considered first (Fig. \ref{fig8_twoperf}A), and then trials with random initial angles are considered (Fig. \ref{fig8_twoperf}D). The system evolves for $T=5$s (500 time units), and then the centroid of the bumps corresponding to $\phi_1$ is read out and compared to the original angle by computing Eq.~(\ref{msedef}). If the bumps merge, both items are represented by the remaining bump. The MSE is similarly computed using the low-dimensional approximation, Eq.~(\ref{twocent}), if the bumps begin sufficiently far apart, otherwise Eq.~(\ref{mergmse}) is used to approximate the MSE. These approximations are compared to the full simulations in Fig. \ref{fig8_twoperf}B,C, for the case in which the initial target angles are $\phi_1 = + x_0$ and $\phi_2 = -x_0$. Merging causes both bumps to have mean position $x=0$, when the initial targets $\pm x_0$ are sufficiently close (Fig. \ref{fig8_twoperf}B). Beyond this boundary, the stored angles repel one another. There is an abrupt transition in the MSE corresponding to this boundary point (Fig. \ref{fig8_twoperf}C). Importantly, the MSE is limited from below at each $x_0$ by the variance a single bump ($\langle \Delta^2 \rangle = D \cdot T$), not interacting with another bump. Thus, even though the peak MSE grows significantly for the case ${\mc A}=10$, it is important to note that the MSE will be significantly smaller at large values of $x_0$, since individual bumps diffuse less for larger values of ${\mc A}$ (Fig. \ref{fig4_onebump}F). 

The lower bound on the MSE produced by a single bump's trajectory is approached when the two bumps are either initiated at the same location ($x_0=0$), or when the bumps are initiated sufficiently far from one another. While there will always be vanishingly small repulsive effects that will tend to push bumps farther apart, we see that even for $x_0 \approx 6$, the MSE appears to approach a lower limit. This is because the long-range interactions between bumps are on the order of $\e^{-12} \approx 6 \times 10^{-6}$, when using the weight function Eq.~(\ref{wizard}). These effects are smaller than the discretization error produced by the spatial mesh of our numerical integration scheme, so we would expect the strength of repulsion to be weaker than the pinning produced by discretizing, as discussed in \cite{guo05}.

Performance on the two-item WM task with random initial targets $\phi_1$ and $\phi_2$ is considered in Fig. \ref{fig8_twoperf}E. Recall variability, represented by the MSE, is greater than what would be predicted by a model that allows distinct slots for each item. Note, there have been efforts to revise the slot model~\citep{zhang08,cowan10}, so that error increases when considering two items versus one item. However, the increases in error arising from neural activity dynamics we observe are much more nuanced than would be possible for previous phenomenological slots or resources models~\citep{zhang08,bays08}. Both items (bumps) are stored in a single network, producing interactions between bumps when items are initially close, which contributes an additional source of variability to the recall. Merging produces a systematic shift in the remembered location of items, as does repelling. The frequency of these interactions grows as the synaptic strength parameter ${\mc A}$ is increased, counteracting the reduction in diffusion also produced by increasing ${\mc A}$. This tradeoff produces a non-monotonic dependence of the MSE on ${\mc A}$ (Fig. \ref{fig8_twoperf}E), so there is an optimal ${\mc A}$ for two-item storage with low-diffusion of bumps and low-probability of bump interaction. This optimum occurs when ${\mc A} \approx 10$, so even though the peak MSE is much larger than for the cases ${\mc A}=1,2$ (Fig. \ref{fig8_twoperf}C), the average MSE is smaller since bumps are less susceptible to stochastic perturbations. Note, the MSE in the interacting bumps model is larger than would be predicted by a slots model that assumes MSE is unchanged as the number of items is increased up to some fixed capacity (Fig. \ref{fig8_twoperf}E).

Our interacting bumps model can account for the item-dependent increase in the variability of recall in two-item WM tasks~\citep{wilken04,bays09}. This arises due to the nonlinear interactions between the bumps, which add to the variability already present due to the dynamic fluctuations in the network. We now examine item-dependent changes in recall variability for tasks with more than two items, showing our analysis extends to the case of multiple interacting bumps.

% Section 4
\section{Multiple interacting bumps}
\label{multiple}

Recent models of multi-item WM focus on uncovering the nature of item-number limitations, as they impact response variability~\citep{ma14}. Phenomenological models can be altered to capture errors that either reflect a finite capacity or the distribution of resources~\citep{zhang08}, but physiologically-inspired models account for the architecture and dynamics of neural circuits underlying WM storage~\citep{bays15}. The work of \citet{edin09,wei12,almeida15} has shown that a recurrent spiking network can support multiple bumps that each individually encode a different item. Our model is a tractable version of these previous studies, allowing us to derive explicit expressions describing limitations of the network. 

Prior to developing effective equations for bump interfaces, we consider the problem of network capacity. This is one way in which our model differs from the standard resource model of WM. Only a finite number of bumps can be stored in the recurrent network, and this upper limit is determined by the choice of the synaptic strength parameter ${\mc A}$. However, we note this upper limit is quite large. We can approximate this limit by again examining a stationary solution problem.

\subsection{Network capacity}

\begin{figure*}
\begin{center} \includegraphics[width=16cm]{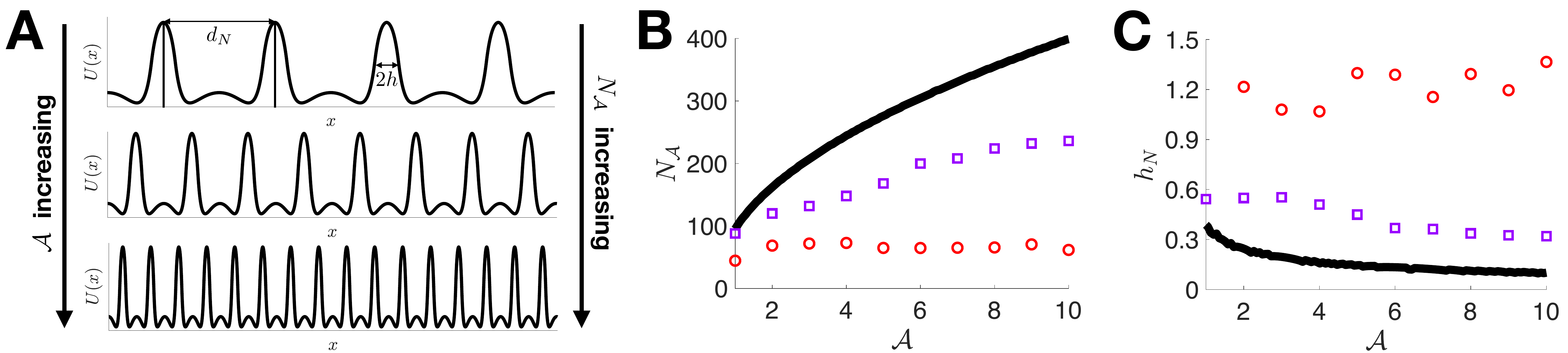}\end{center}
\caption{Capacity of the network, Eq.~(\ref{nfield}), is bounded by the number of stable bumps $N_{\mc A}$ that can be instantiated in the periodic domain $x \in [-180,180)$. {\bf A}. The capacity tends to increase with synaptic strength $\A$, since stronger synapses lead to narrower bumps in multibump solutions. {\bf B}. The maximal number $N_{\mc A}$ of bumps that can be packed into the domain increases with ${\mc A}$. Periodic solutions have spacing $d_N = 360/N$ between bump centroids, and the same half-width $h_N$ for each bump. Solid line shows result from theory Eq.~(\ref{Nbumpcons}), squares are results from deterministic simulations, and circles are from stochastic simulations (See main text). Note for stochastic networks, the capacity is fairly flat across ${\mc A}$. {\bf C}. The bump half-width $h_N$ associated with the maximal capacity solution narrows for increasing synaptic strength ${\mc A}$ owing to the strong recurrent inhibition for networks with large ${\mc A}$. Threshold $\theta = 0.25$.}
\label{fig9_multibump}
\end{figure*}

We frame the problem of identifying network capacity by attempting to identify multi-bump stationary solutions to Eq.~(\ref{nfield}) in the absence of noise ($Z \equiv 0$). Finite multi-bump solutions are not stable in the limit $L \to \infty$~\citep{laing03b}, since multiple active regions exert a repulsive drift on one another. If bumps are spaced evenly around the domain, the conformation is stable since the repulsive forces acting on each bump from either direction balance. Thus, stable multi-bump solutions constitute a periodic pattern that wraps around the domain. One question is just how the minimal period of this pattern changes as the synaptic strength $\A$ is changed. Since $\A$ increases the width of single bump solutions, one might expect the capacity to decrease as $\A$ is increased. We demonstrate in fact that the capacity of the network grows as the synaptic strength ${\mc A}$ is increased (Fig. \ref{fig9_multibump}A).

Network capacity can be bounded by examining the upper limit on the number of possible bumps in a periodic solution to Eq.~(\ref{nfield}). These numbers will tend to be much larger than those imposed by a slots model of WM capacity~\citep{zhang08}, so our model will still behave approximately as a resource model since the capacity $N_{\mc A}$ is quite high. The capacity can be estimated by studying the existence of multi-bump solutions, comprised of multiple stationary active regions of the same width, spaced an even distance apart. For example, a two-bump solution with centroids at $x = \pm 90$ has the form
\begin{align*}
U(x) &= \int_{-90-h_2}^{-90+h_2} w(x-y) \d y + \int_{90-h_2}^{90+h_2} w(x-y) \d y.
\end{align*}
As in the case of single bumps, there is one unknown, which is the half-width of each bump $h_2$. Self-consistency of the threshold conditions $U(-90 \pm h_2) = U(90 \pm h_2) = \theta$ yields an implicit equation
\begin{align}
\theta &= \int_0^{2h_2} w(y) \d y + \int_{180-2h_2}^{180} w(y) \d y,   \label{twobumpthresh}
\end{align}
which follows from the periodicity of the weight function ($w(180+\alpha) = w(180-\alpha)$). Computing integrals in Eq.~(\ref{twobumpthresh}) for the weight function Eq.~(\ref{wizard}), we find the implicit equation for the half width $h_2$ is
\begin{align}
\theta = 2 {\mc A} \left( h_2 \e^{-2h_2} + \e^{-180} \left[ 90 - (90-h_2)\e^{2h_2} \right] \right),  \label{twobumph}
\end{align}
so if Eq.~(\ref{twobumph}) has a solution we expect the network with synaptic strength ${\mc A}$ to have capacity of at least two items. Note $\e^{-180} \approx 6.17 \times 10^{-79}$ is extremely small, so networks in which single bumps exist will likely also possess two bump solutions, since Eq.~(\ref{twobumph}) is a very mild perturbation of Eq.~(\ref{widform}). This approach can be generalized to the case of more than two bumps ($N>2$), yielding analogous equations to Eq.~(\ref{twobumph}) for the associate bump half-width.

Solutions with $N$ distinct bumps have a regular spacing between each of the bump, forming a periodic pattern that tiles the domain. The spacing between the centroid of each bump is computed by partitioning the domain by $N$, $d_N = 360/N$. We can write an $N$-bump solution in the form
\begin{align*}
U(x) &= \sum_{j=1}^N \int_{c_j -h_N}^{c_j + h_N} w(x-y) \d y
\end{align*}
where $c_j = -180+ (j + 1/2)d_N$ is the location of the centroid of the $j^{\text{th}}$ bump. Threshold conditions $U(c_j \pm h_N) = \theta$ for $j=1,...,N$ yield the implicit equation for the half-width
\begin{align}
\theta = \sum_{j=0}^{N-1} \int_{j d_N}^{jd_N + 2h_N} w(y) \d y. \label{multithresh}
\end{align}
%which can be rewritten using periodicity ($w(180+\alpha) = w(180-\alpha)$)
%\begin{subequations} \label{multithresh}
%\begin{align}
%\theta &= \int_0^{2h_N} w(y) \d y + \sum_{j=1}^{(N-1)/2} \int_{jd_N - 2h_N}^{jd_N + 2h_N} w(y) \d y, \hspace{40mm} N \ \text{odd}, \\
%\theta &= \int_0^{2h_N} w(y) \d y + \sum_{j=1}^{(N-2)/2} \int_{jd_N - 2h_N}^{jd_N + 2h_N} w(y) \d y + \int_{180-2h_N}^{180} w(y) \d y, \hspace{13mm} N \ \text{even}.
%\end{align}
%\end{subequations}
For $N$ large, Eq.~(\ref{multithresh}) will contain many very small terms corresponding to the interactions between distant bumps. It is easier to express the sum of integrals in Eq.~(\ref{multithresh}) by approximating the $N$-bump solution on $x \in [-180,180)$ with an infinite-bump solution on $x \in (- \infty, \infty)$. Placing the core bump at $x=0$, and other bumps with centroids at $j d_N$ for $j$ a nonzero integer, yields the implicit equation
\begin{align*}
\theta = \sum_{j=-\infty}^{\infty} \int_{j d_N}^{jd_N + 2h_N} w(y) \d y,
\end{align*}
which we can integrate directly using the weight function Eq.~(\ref{wizard}), resulting in the following equation that must be solvable for a neural field on $ x \in (-\infty, \infty)$ to have periodic bump solutions spaced $d_N$ apart:
\begin{align}
\theta &= {\mc A} \left( 2 h_N \e^{- 2 h_N} + {\mc S}(h_N, d_N) \right),  \label{Nbumpcons}
\end{align}
where
\begin{align*}
{\mc S}(h, d) &= \sum_{j=1}^{\infty}  \e^{-jd} \left[ 4h \cosh (2h) - 2 j d \sinh (2h) \right],  \\
%&= 4h \cosh (2h) \sum_{j=1}^{\infty} \e^{-jd} - 2d \sinh(2h) \sum_{j=1}^{\infty} j \e^{-jd}, \\
&= \frac{4h \cosh (2h) }{\e^d - 1} - \frac{2d \e^{d} \sinh(2h)}{( \e^d - 1)^2}.
\end{align*}
%The integrals in Eq.~(\ref{multithresh}) can be computed directly in the case of the weight function Eq.~(\ref{wizard}), resulting in the following equation that must be solvable for $N$ bumps to exist in the neural field
%\begin{subequations} \label{multibsolve}
%\begin{align}
%\theta &= {\mc A} \left( 2 h_N \e^{-2h_N} + {\mc S}((N-1)/2,N) \right), \hspace{20mm} N \ \text{odd}, \\
%\theta &= {\mc A} \left( 2 h_N \e^{-2h_N}  + \e^{-180} \left[ 180 - (180-2h_N)\e^{2h_N} \right] + {\mc S}((N-2)/2,N) \right), \hspace{3mm} N \ \text{even},
%\end{align}
%\end{subequations}
%where we define
%\begin{align*}
%{\mc S}(M,N) = \sum_{j=1}^M \e^{-j d_N} \left[ (j d_N +2h_N) \e^{- 2h_N} - (j d_N - 2h) \e^{2h_N} \right].
%\end{align*}
We also require $h_N \in [0, d_N/2)$, since the active region of each bump cannot overlap with another. Thus, the capacity $N_{\mc A}$ is approximated by the maximum value $N$ for which Eq.~(\ref{Nbumpcons}) possesses a solution. We plot $N_{\mc A}$ as a function of ${\mc A}$ in Fig. \ref{fig9_multibump}B (solid line), showing our analytically determined capacity is an increasing function of synaptic strength. This may seem surprising, since one might expect that bumps will be wider as ${\mc A}$ grows. However, this is only true when bump interactions are not considered. Many bumps will tend to interact through strong inhibition, which narrows them, decreasing $h_N$ (Fig. \ref{fig9_multibump}C), but they will still be sustained by the strong recurrent excitation generated by increasing ${\mc A}$.

Due to both the infinite domain approximation in Eq.~(\ref{Nbumpcons}) and inevitable truncation errors in our numerical root-finding scheme, we find that the maximal number of bumps predicted by our theory overestimates what we find in deterministic simulations (circles in Fig. \ref{fig9_multibump}B,C). Recall, we are identifying roots of a transcendental equation, which can be quite sensitive to truncation errors. Thus, in practice we would certainly not expect the capacity predicted by our theory to be obtained in simulations of Eq.~(\ref{nfield}) with the addition of stochastic forcing. While calculations using Eq.~(\ref{Nbumpcons}) provide a clean method for estimating the upper bound on bump number in the network, in practice, we expect these solutions to be sensitive to perturbations arising in a numerical integration scheme (See for example discussion in \cite{guo05}).

Thus, we performed a coarser estimate of the network capacity by using a simple numerical simulation method, to find how many bumps can be packed into the domain $x \in [-180,180)$ for each ${\mc A}$. We ran deterministic numerical simulations of Eq.~(\ref{nfield}) in the absence of noise with the initial condition $u(x,0) = \sin (N_{\mc A} \pi x/180)$, allowing them to equilibrate after long time ($t\to 5$s). At this point, we counted the number of bumps remaining, and computed their half-width, plotting the result in comparison to our analysis in Fig. \ref{fig9_multibump}B,C (squares). In fact, this method provides a considerably lower bound on the estimate, but the trend of the maximum bump number increasing with $\A$ is still present.

\begin{figure*}
\begin{center} \includegraphics[width=15cm]{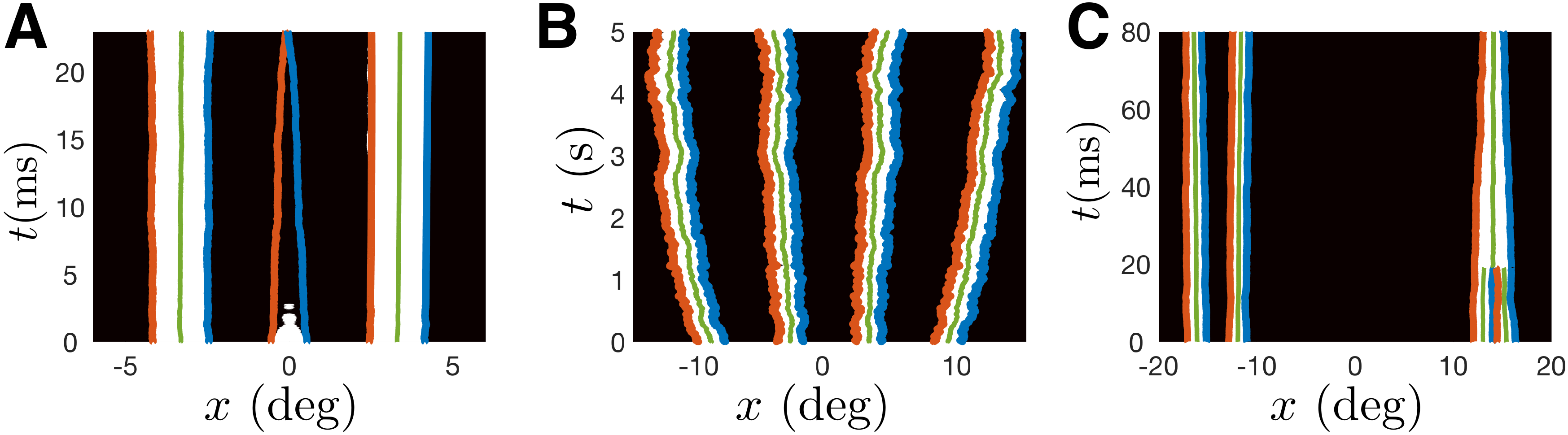} \end{center}
\caption{Categories of bump interactions that induce WM errors. {\bf A}. Annihilation occurs when a bump is closely flanked by two others, leading to extinction of the inner bump. {\bf B}. Repulsion drives bumps away from one another, due to the recurrent inhibition acting at long distances. {\bf C}. Bumps will merge if two begin close and are not in the vicinity of other bumps. Interface equations, Eq.~(\ref{multiface}), track the dynamics of full numerical simulations of Eq.~(\ref{nfield}). Parameters are $\ep = 0.03$ and $\theta = 0.25$.}
\label{fig10_multisim}
\end{figure*}

To estimate capacity from stochastic simulations, we used the steady state of deterministic simulations as an initial condition for $10^4$ Monte Carlo simulations of Eq.~(\ref{nfield}) for $t=5$s, using parameters from Fig. \ref{fig4_onebump}. We then averaged the number of bumps at the end of all simulations. In this case, we found the capacity is fairly insensitive to ${\mc A}$ (Fig. \ref{fig9_multibump}B), but typically in the range of 60--70 bumps, well above the single digit range predicted in \cite{edin09,wei12}. A major reason for this is that these previous models assumed a wider footprint of inhibitory connectivity, whereas our model assumes inhibition with a similar spatial scale to excitation (See also \cite{lim14,rosenbaum17}). 

%We now examine dynamic interactions of bumps in numerical simulations of Eq.~(\ref{nfield}) for the case of a more than two bumps, and derive a low-dimensional system for the interface dynamics.

\subsection{Interface equations}

We extend the interface equations derived for the interactions of two bumps to account for interactions between an arbitrary number of bumps in the network. As in the case of two bumps, if multiple bumps' initial active regions overlap, they will merge. This sets an upper bound on the capacity of the network (as in Fig. \ref{fig9_multibump}). Our analysis proceeds now by projecting the dynamics of bumps in the network to equations that simply track bump interfaces.

The active region in this case is given by the union of $N$ finite intervals, which we assume to be disjoint, $A(t) = \cup_{j=1}^N \left[ a_j(t), b_j(t) \right]$. Note, in the case that bumps' active regions overlap, we simply redefine a set of fewer than $N$ finite intervals, with some corresponding to the merged bumps. In the case of $N$ distinct active regions, Eq.~(\ref{nfield}) becomes
\begin{align*}
\d u(x,t) &= \left[ -u(x,t) + \sum_{j=1}^{N} \int_{a_j(t)}^{b_j(t)} w(x-y) \d y \right] \d t \\
& \hspace{22mm} + \sqrt{\ep \cdot |u(x,t)|} \d Z(x,t).
\end{align*}
The dynamic threshold equations are then given
\begin{align}
u(a_j(t),t) = u(b_j(t),t) = \theta, \hspace{5mm} j=1,...,N.  \label{dthreshn}
\end{align}
As before, we differentiate Eq.~(\ref{dthreshn}), rewrite the corresponding integrals using Eq.~(\ref{wanti}), and assume the gradients at the threshold are static and approximated by the stationary bump gradients ($\bar{\alpha} = U_0'(-h)$). We can approximate the interface dynamics by the following stochastic system:
\begin{subequations} \label{multiface}
\begin{align}
\d a_j &= \frac{1}{\bar{\alpha}} \left( \left[ \theta - \sum_{k=1}^N (W(a_j-a_k)-W(a_j-b_k)) \right] \d t \right. \nonumber \\
& \hspace{26mm} \left. - \sqrt{\ep \cdot \theta} \hspace{0.5mm} \d Z(a_j,t) \right), \\
\d b_j &= - \frac{1}{\bar{\alpha}} \left( \left[ \theta -  \sum_{k=1}^N (W(b_j-a_k)-W(b_j-b_k)) \right] \d t \right. \nonumber \\
& \hspace{26mm} \left. - \sqrt{\ep \cdot \theta} \hspace{0.5mm} \d Z(b_j,t) \right), 
\end{align}
\end{subequations}
for $j=1,...,N$. Furthermore, the dynamics of the centroid of each bump $j$ can be tracked by using the change of variables, $\Delta_j = (a_j+b_j)/2$, along with the approximations $a_j \approx \Delta_j -h$ and $b_j \approx \Delta_j + h$, yielding the system of $N$ stochastic differential equations
\begin{align}
\d \Delta_j &= \frac{1}{\bar{\alpha}} \left( \sum_{k \neq j} J(\Delta_k - \Delta_j) \d t+ \frac{\sqrt{\ep \cdot \theta}}{2} \d {\mc Z}(\Delta_j,t) \right),  \label{multicent}
\end{align}
for $j=1,...,N$, where $J(\Delta)$ and $\d {\mc Z}(\Delta_j,t)$ are defined as in Eq.~(\ref{twocent}).  While we could employ the low-dimensional Eq.~(\ref{multicent}) to approximate how the network, Eq.~(\ref{nfield}), performs on multi-item WM tasks, we opt to preserve the interface information in Eq.~(\ref{multiface}). Interactions between multiple bumps are much better captured when the width of bumps is also considered along with their position. Truncating to only consider the centroid ignores width perturbations, which for example ignores bump annihilation events.

\begin{figure*}
\begin{center} \includegraphics[width=15.5cm]{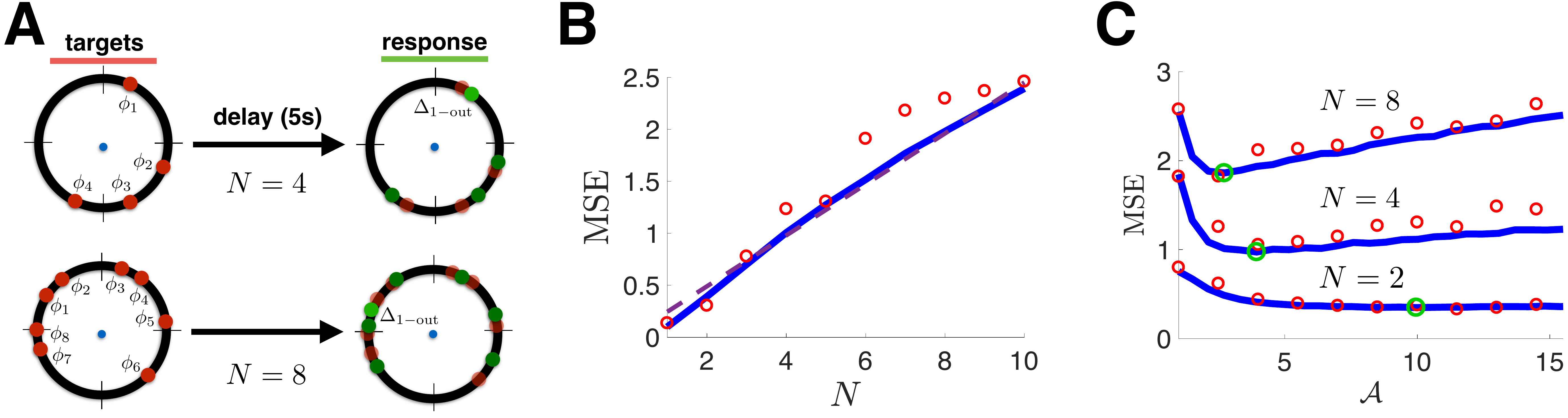} \end{center}
\caption{Recall error in multi-item WM task. {\bf A}. There are $N$ target items $\{ \phi_1, \phi_2, ..., \phi_N\}$ whose locations are chosen uniformly randomly around $x\in [-180,180)$. Mean squared error (MSE) is computed by comparing the output angle $\Delta_{1 - \text{out}}$ of a randomly chosen input $\phi_1$ via Eq.~(\ref{msedef}). {\bf B}. MSE increases with the number of items $N$, computed from Eq.~(\ref{multiface}), (solid line), agreeing with numerical simulations of Eq.~(\ref{nfield}), (circles). Trend is fit by a linear function (dashed line). We fix $\A = 5$.  {\bf C}. MSE is nonmonotonic in synaptic strength ${\mc A}$, owing to the tradeoff between fluctuation-driven errors (at small ${\mc A}$), and bump interaction errors (at large ${\mc A}$). Note, the optimal ${\mc A}$ (large circles) decreases as $N$ is increased.}
\label{fig11_multiperf}
\end{figure*}

We demonstrate the efficacy of the low-dimensional description, Eq.~(\ref{multiface}) in capturing the dynamics of the bump interfaces in Fig. \ref{fig10_multisim}. One category of dynamics that is much more common in the case of multiple bumps is annihilation of a bump by two neighboring bumps on either side (Fig. \ref{fig10_multisim}A). Notice, the middle bump does not merge with the bump on the left or right, but is extinguished by their combined recurrent inhibition. This is captured reasonably well by our interface Eq.~(\ref{multiface}). Note, annihilation events can also occur in the case of two bumps being instantiated close to one another. The combination of noise and lateral inhibition can lead to one bump's extinction without it merging with the neighboring bump. Bump edges fluctuate in response to noise, and multiple bumps can collectively repel each other when they are not too close (Fig. \ref{fig10_multisim}B). Merging occurs when two bumps begin close to one another, but are not in the vicinity of other bumps (Fig. \ref{fig10_multisim}C).

\subsection{Performance}

Estimation errors for WM tasks involving an arbitrary number of targets $N$, $\{ \phi_1, \phi_2, ..., \phi_N\}$, still mostly arise from merging, repulsion, or diffusion. However, recurrent inhibition from multiple active regions can also result in annihilation of bumps, so the activity is extinguished separately from a neighboring bump (as shown in Fig. \ref{fig10_multisim}A). These errors combine to shape the response variability, measured by the MSE in Eq.~(\ref{msedef}), as a function of the item number $N$. Our results demonstrate much more consistency with a resource model of multi-item WM. In particular, we see that the response variability for individual items increases with the number of items stored in memory, starting with the difference between $N=1$ and $N=2$ discussed in the previous section (Fig. \ref{fig8_twoperf}E).

We have to make specific choices about how multiple bumps encode the multiple target items that initially instantiate them. As in the case of two-items, we ignore probabilistic effects that could further contribute to the error beyond that captured by the stochastic dynamics of the bumps~\citep{bays09}. Lapses and swaps are not considered in our model. When two bumps merge with one another, or one bump is annihilated, the items associated with vanishing bumps are then associated to the closest remaining bump. Handling bump annihilation events parsimoniously singles out the contributions to the error engendered solely by the neural circuit dynamics described by our recurrent network. 

The MSE of a single item is computed using the formula, Eq.~(\ref{msedef}), as before. Recall, we assume that a subject is probed about a single item in WM, as this is the protocol typical used to measure behavioral errors~\citep{ma14}. Furthermore, this allows us to significantly reduce the space of dynamics that we must track in each numerical simulation. Rather than having to track the locations of all bumps, it is sufficient to only follow the dynamics of bumps in the vicinity of the bump originating from the item of interest. We examine performance when target angles $\{ \phi_1, \phi_2, ..., \phi_N\}$ are randomly placed about the domain $x \in [-180,180)$ (Fig. \ref{fig11_multiperf}A). First, we study the scaling of recall variability, measured by MSE, as a function of the number of items stored $N$. As expected by our two-bump performance results, we find that the MSE grows steadily with $N$ (Fig. \ref{fig11_multiperf}B). This is suggestive of the resource model of WM, since the increase occurs across all $N$. Note, the simulations of the interface equations, Eq.~(\ref{multiface}), (solid line) agree with with simulations of the full model, Eq.~(\ref{nfield}), even though the interface equations are considerably faster to numerically solve. In addition, we find the MSE trend is well fit by a linear function of $N$. Next, we study the effect of varying the synaptic strength ${\mc A}$ for different item number counts. Extending our findings for two-item memory, we see there is an optimal ${\mc A}$ that minimizes the MSE, and this optimum decreases as $N$ is increased (Fig. \ref{fig11_multiperf}C). Thus, as more items must be stored in the network, it is advantageous to have networks with narrower bumps to decrease the probability of bump interactions.

\section{Extensions to multiple stimulus features}
\label{multfeat}

Thus far, we have studied a network that represents multiple items in WM as bumps within a single-layer neural field. However, WM tasks typically employ stimuli with multiple features (e.g., color and orientation), which likely are represented by populations of neurons with tuning that varies along multiple feature dimensions~\citep{mante13,schneegans17}. In light of this, we now consider an extension of Eq.~(\ref{nfield}) to a multilayer neural field model with spatially-organized interlaminar connectivity. Each distinct layer represents a different value of a quantized stimulus feature.
For example, the discrete stimulus feature may be color, and the continuous variable, represented by position within each layer, could encode orientation (as represented in Figs \ref{fig8_twoperf}A,D and \ref{fig11_multiperf}A).
When multi-item WM is represented by the multilayer neural field, interaction-based errors decrease since stimuli of different colors interact less.

\begin{figure*}
\begin{center} \includegraphics[width=14cm]{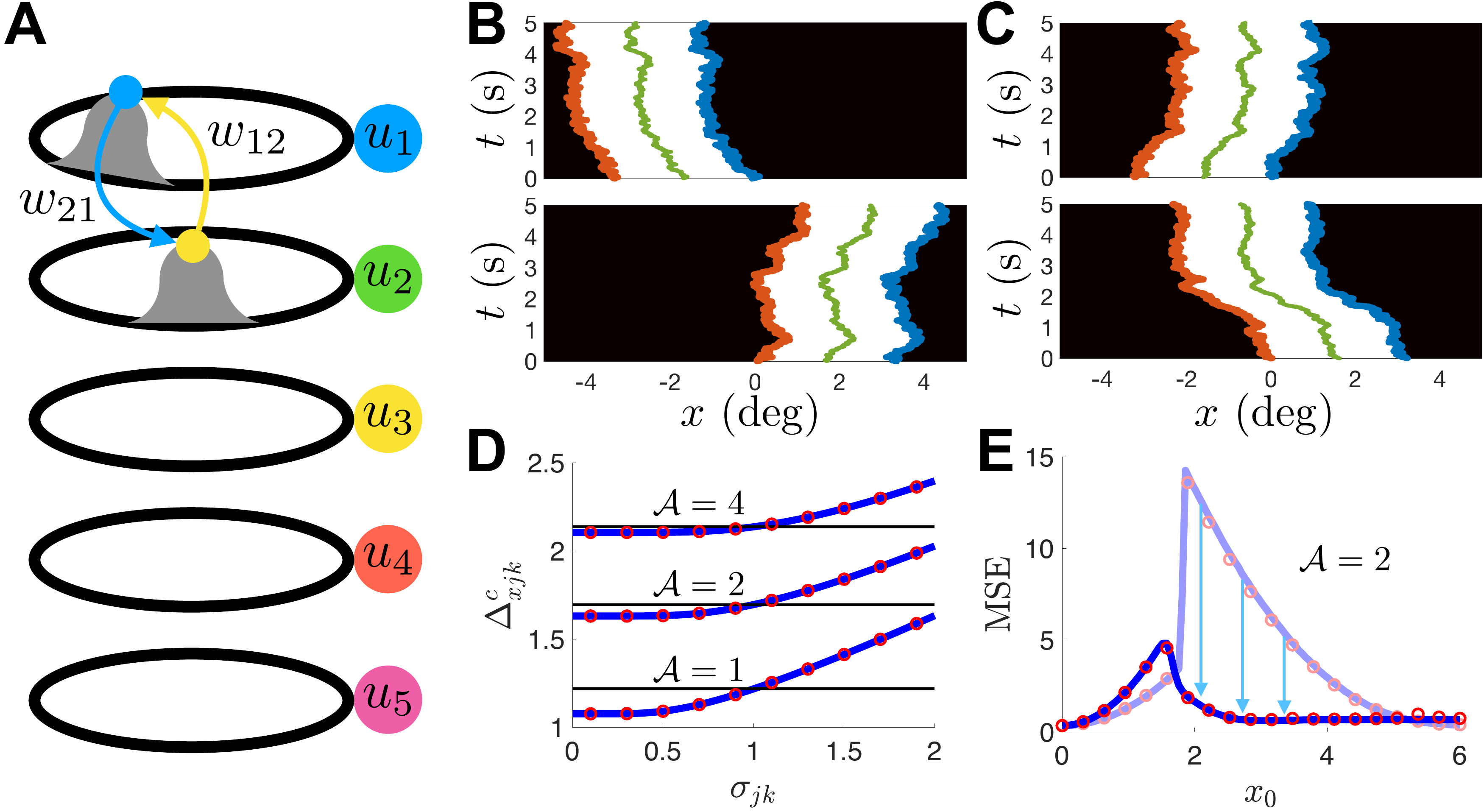} \end{center}
\vspace{-4mm}
\caption{Interaction-related errors are reduced in a multilayer network model, Eq.~(\ref{mfield}). {\bf A}. Recurrent network is now comprised of multiple rings with both within layer connectivity $w_{jj}(x-y)$ and interlayer connectivity $w_{jk}(x-y)$, as described by Eq.~(\ref{interexp}). The centroid location $x$ of a bump within a layer represents an encoded orientation, whereas the layer in which it resides encodes color. Bumps in distinct layers communicate via weak interlayer connectivity. {\bf B}.~Bumps initiated in two distinct layers (1 and 2) repel one another when initiated far enough apart in the $x$-dimension (at $\pm x_0 = \pm 1.8$ here). {\bf C}. Bumps initiated in two distinct layers (1 and 2) are attracted to one another when initiated close enough together in the $x$-dimension (at $\pm x_0 = \pm 1.6$ here). {\bf D}. The critical boundary $\Delta_{xjk}^c$ between collocation and repulsion is determined by Eq.~(\ref{crittwolay}), (solid curves), and agrees with direct simulations (circles). Note, the minimal distance for bumps to repel one another increases as the width of interlaminar connections $\sigma_{jk}=\sigma_{kj}$ is increased. {\bf E}. MSE as a function of $x_0$ computed from Eq.~(\ref{msedef}) is maximized near $\Delta_{xjk}^c$, where repulsion is strongest, in the multilayer neural field, Eq.~(\ref{mfield}), for bumps initiated in different layers (1 and 2). Low-dimensional approximation, Eq.~(\ref{laycent}), (dark blue line) agrees well with simulations of full model Eq.~(\ref{mfield}) (dark red circles). Compare with the MSE in the single-layer neural field, Eq.~(\ref{nfield}), which is appreciably larger (light line and circles). Unless stated otherwise, parameters are $\theta = 0.25$, ${\mc A} = 2$, ${\mc A}_{12} = {\mc A}_{21} = 0.1$, $\sigma_{12}=\sigma_{21}=0.5$, $\ep = 0.03$, $w_c = 25 \pi /180$, and $\bar{C}_{jk} = \delta_{jk}$.}
\vspace{-4mm}
\label{fig12_multilay}
\end{figure*}

The multilayer neural field model we analyze builds on a model introduced in \cite{kilpatrick13c}, which represents WM for a single-item using multiple bumps in distinct but coupled network layers. Here, we consider a variation of this model that allows for the representation of multiple items, each with their own bump, possibly in distinct layers of the network (Fig. \ref{fig12_multilay}A). Activity in this neural field model is now described by the following set of evolution equations:
\begin{align}
\d u_j(x,t) &= \left[ - u_j(x,t) + \sum_{k=1}^{M} w_{jk}*H(u_k - \theta) \right] \d t \nonumber \\
& \hspace{22mm} + \sqrt{\ep \cdot |u_j|} \d Z_j,  \label{mfield}
\end{align}
where $j=1,...,M$ are the layers of the network. Each distinct layer represents a different color value, whereas spatial locations $x$ correspond to the preferred stimulus orientation of neurons there. Synaptic connectivity within and between layers is described by layer-dependent weight function
\begin{align}
w_{jk}(x) = {\mc A}_{jk} \left( 1- \frac{|x|_L}{\sigma_{jk}} \right) \e^{-|x|_L/\sigma_{jk}},  \label{interexp}
\end{align}
so ${\mc A}_{jk}$ and $\sigma_{jk}$ scale the amplitude and width of synaptic connectivity. We chose to normalize connectivity within layers, ${\mc A}_{jj} := {\mc A}$, and take weaker connectivity between layers $0 \leq {\mc W}_{jk} < {\mc A}$ ($j \neq k$). In addition, we normalize the width of connectivity within layers $\sigma_{jj} :=1$ and take narrow connectivity between layers, $0 \leq \sigma_{jk}<1$. The antiderivative of Eq.~(\ref{interexp}) is then given
\begin{align}
W_{jk}(x) = \int_0^x w_{jk} (y) \d y = {\mc A}_{jk} x \e^{-|x|/ \sigma_{jk}}, \label{layanti}
\end{align}
and $W_{jj}(x) = W(x)$ from Eq.~(\ref{wanti}). Noise to each layer is weak and multiplicative $\sqrt{\ep \cdot |u_j(x,t)|} \d Z_j(x,t)$, where we define spatially-dependent Wiener processes such that $\langle \d Z_j(x,t) \rangle = 0$ and $\langle \d Z_j(x,t) \d Z_k(y,s) \rangle = C_{jk} (x-y) \delta (t-s) \d t \d s$ for $j,k=1,...,M$. In numerical simulations, we consider independent cosine correlated noise to each layer $C_{jj}(x) = \cos (\omega_c x)$, where $\omega_c = c \pi / L$, so that $C_{jk}(x) \equiv 0$ ($j \neq k$), but see \cite{kilpatrick13c} for more details on the impact of noise correlations between layers.

We can extend our analysis of Eq.~(\ref{nfield}) to consider interactions of multiple bumps in Eq.~(\ref{mfield}).  Our reduction proceeds by projecting the dynamics of bumps in multiple layers to equations tracking interfaces of each bump in its respective layer. Note that now bumps can only merge if they begin in the same layer $j$. If bump centroids are initially close in the $x$-dimension but reside in different layers, the bumps will colocate to roughly the same $x$ location in their respective layers (Fig. \ref{fig12_multilay}B). Repulsion between layers can occur as well (Fig. \ref{fig12_multilay}C). For $N$ bumps, the active region is given by $N$ finite intervals, assumed to be disjoint:
\begin{align*}
A(t) = \cup_{n=1}^N \left[ a_n(t), b_n(t) \right] \times \{ l(n) \},
\end{align*}
where $\{ l(n) \}$ denotes the layer in which bump $n$ resides. As before, if bumps' active regions overlap, we redefine a smaller set of intervals. In the case of $N$ active regions, Eq.~(\ref{mfield}) becomes
\begin{align*}
\d u_j &= \left[ - u_j + \sum_{n=1}^N \int_{a_n(t)}^{b_n(t)} w_{jl(n)}(x-y) \d y \right] \d t \\
&\hspace{3mm}  + \sqrt{\ep \cdot |u_j(x,t)|} \d Z_j(x,t).
\end{align*}
The dynamic threshold equations are then given by
\begin{align}
u_{l(n)}(a_n(t),t) = u_{l(n)}(b_n(t),t) = \theta,    \label{mlaythresh}
\end{align}
for $n=1,...,N$. Differentiating Eq.~(\ref{mlaythresh}), rewriting the integrals using Eq.~(\ref{wanti}), and using the static gradient approximation, we can write the interface dynamics as
\begin{subequations} \label{mlayiface}
\begin{align}
& \d a_n = \frac{1}{\bar{\alpha}} \left( \left[ \theta - \sum_{m=1}^N (W_{l(n)l(m)}(a_n - a_m) \right. \right. \\
& \left. \left. - W_{l(n)l(m)}(a_n- b_m)) \right] \d t - \sqrt{\ep \cdot \theta} \d Z_n(a_n,t) \right), \nonumber \\
&\d b_n = - \frac{1}{\bar{\alpha}} \left( \left[ \theta - \sum_{m=1}^N (W_{l(n)l(m)}(b_n - a_m)  \right. \right. \\
& \left. \left. - W_{l(n)l(m)} (b_n - b_m)) \right] \d t - \sqrt{\ep \cdot \theta} \d Z_n(b_m,t) \right), \nonumber
\end{align}
\end{subequations}
for $j=1,...,N$, where the layer-dependence of the bump interactions is expressed via the modulation of the interlaminar weight antiderivatives $W_{l(n)l(m)}$ given by Eq.~(\ref{layanti}).

We now demonstrate how this multilayer network structure can lead to an effective reduction in bump interactions, even in the noise-free system ($Z_j \equiv 0$, $\forall j = 1,...,M$). For simplicity, we focus on the case of two symmetrically-placed bumps in two distinct layers
\[
A(0) = [-b_0,-a_0] \times \{j\} \cup [a_0,b_0] \times \{ k \},
\]
of a symmetric network ($W_{jk}(x) \equiv W_{kj}(x)$, $\forall j,k$). Therefore, we can write Eq.~(\ref{mlayiface}) as 
\begin{subequations} \label{twofacelay}
\begin{align}
a_1'(t) &= \frac{1}{\bar{\alpha}} \left[ \theta - W(b_1-a_1) - W_{jk}(b_2- a_1) \right. \nonumber \\
& \hspace{25mm} \left. +W_{jk}(a_2 - a_1) \right], \\
b_1'(t) &= - \frac{1}{\bar{\alpha}} \left[ \theta - W(b_1-a_1) -W_{jk} (b_2 - b_1) \right. \nonumber \\
& \hspace{25mm} \left. + W_{jk}(a_2 - b_1)  \right], \\
a_2'(t) &= \frac{1}{\bar{\alpha}} \left[ \theta - W(b_2 -a_2) - W_{kj}(a_2-a_1) \right. \nonumber \\
& \hspace{25mm} \left.  + W_{kj}(b_2-a_1) \right], \\
b_2'(t) &= -\frac{1}{\bar{\alpha}} \left[ \theta - W(b_2-a_2) - W_{kj}(b_2-a_1) \right. \nonumber \\
& \hspace{25mm} \left.  +W_{kj}(b_2-b_1) \right],
\end{align}
\end{subequations}
so by symmetry, $a(t) : = a_2(t) = -b_1(t)$ and $b(t):=b_2(t) = -a_1(t)$, and Eq.~(\ref{twofacelay}) reduces to
\begin{subequations} \label{twosymlay}
\begin{align}
a'(t) &= \frac{1}{\bar{\alpha}} \left[ \theta - W(b(t) - a(t)) +W_c(2a(t)) \right. \nonumber \\
& \hspace{27mm} \left.  -  W_c(a(t)+b(t)) \right], \\
b'(t) &= \frac{1}{\bar{\alpha}} \left[ W(b(t)-a(t)) - \theta + W_c(2b(t)) \right. \nonumber \\
& \hspace{27mm} \left.  - W_c(a(t)+b(t)) \right],
\end{align}
\end{subequations}
where $W_c(x) := W_{jk}(x) \equiv W_{kj}(x)$. Note, Eq.~(\ref{twosymlay}) is an approximation of the evolution of the interfaces, since we use a static approximation for the interface gradients. However, the fixed points of Eq.~(\ref{twosymlay}) and the exact system should be the same. As in the single-layer case, there is a critical distance $\Delta_{xjk}^c$ which divides solutions that drift apart ($\Delta_x > \Delta_{xjk}^c$) from those tend towards collocation ($\Delta_{x} < \Delta_{xjk}^c$), given that one bump is in layer $j$ and the other in $k$.  An implicit analytical expression for $\Delta_{xjk}^c$ can be determined by looking for the equilibria of Eq.~(\ref{twosymlay}). As before, the critical curve $(a^c,b^c)$ is determined by the zeros of the right hand side of the $a'(t)$ equation, yielding
\begin{align}
\theta = W_c(b^c-a^c) + W_c(a^c+b^c) - W_c(2a^c), \label{crittwolay}
\end{align}
a curve dividing initial conditions $(a(0),b(0))$ into those that lead to collocation and those that repel. As before, we can assume $b^c - a^c = 2h$, and find the critical curve in terms of the stationary bump half-width $h$ and centroid $\Delta_{xjk}^c= (a^c+b^c)/2$, yielding $W_c(2(\Delta_{xjk}^c - h)) = W_c(2\Delta_{xjk}^c)$. Defining ${\mc A}_c : = {\mc A}_{jk} = {\mc A}_{kj}$ and $\sigma_c : = \sigma_{jk} = \sigma_{kj}$, we can thus apply Eq.~(\ref{layanti}) and solve for $\Delta_{xjk}^c = h/(1-\e^{-2h/\sigma_{jk}})$. Clearly, $\Delta_{xjk}^c$ decreases with $\sigma_{jk}$. Thus, since $0 \leq \sigma_{jk}< 1$ ($j \neq k$), we expect the critical distance $\Delta_{xjk}^c$ to be narrower here than in the single layer case (Fig. \ref{fig12_multilay}D). Interestingly, there is no dependence of $\Delta_{xjk}^c$ on the amplitude ${\mc A}_{jk}$ of interlaminar connectivity, although this will affect the rate at which the bump collocate. Since bumps are in different layers ($j \neq k$), both of them will still remain after a collocation event and will be then centered at the same location in their respective layers (See also \cite{folias11}).

To consider the combination of bump interactions and stochastic forcing on the dynamics of bumps in the multilayer network, Eq.~(\ref{mfield}), we also derive a reduction of Eq.~(\ref{mlayiface}), which tracks the centroid of two bumps. For the purposes of comparison with the single-layer model, we gain significant insight by comparing performance on two-item WM tasks. Starting with Eq.~(\ref{mlayiface}) assuming $N=2$, we assume the width of each bump remains constant, so $a_j = \Delta_j - h$ and $b_j = \Delta_j + h$ for $j=1,2$. As before, we assume symmetric interlaminar connectivity, $W_c(x) := W_{12}(x) \equiv W_{21}(x)$, and we also assume noise is symmetric and independent. The resulting equations for the bump centroids are
\begin{subequations} \label{laycent}
\begin{align}
\d \Delta_1 &= \frac{1}{\bar{\alpha}} \left( J_{c}(\Delta_2 - \Delta_1) \d t + \frac{\sqrt{\ep \cdot \theta}}{2} \d {\mc Z}_1(\Delta_1,t) \right), \\
\d \Delta_2 &= \frac{1}{\bar{\alpha}} \left( J_{c}(\Delta_1 - \Delta_2) \d t + \frac{\sqrt{\ep \cdot \theta}}{2} \d {\mc Z}_2(\Delta_2,t) \right),
\end{align}
\end{subequations}
where
\begin{align*}
&J_{c}(\Delta) = \frac{1}{2} \left[ 2 W_{c} (\Delta) - W_{c}(\Delta - 2h) - W_{c}(\Delta+2h) \right], \\
& \d {\mc Z}_j(\Delta_j,t) = \d Z_j(\Delta_j +h,t) - \d Z_j(\Delta_j - h,t).
\end{align*}
Our low-dimensional approximation, Eq.~(\ref{laycent}), for the evolution of the centroids can be used to compute the estimation errors of the network Eq.~(\ref{mfield}) in the case of two target locations, $\phi_1$ and $\phi_2$. Similar to the single-layer network, Eq.~(\ref{nfield}), errors arise due to merging, repulsion, and diffusion, but the effects of interactions between bumps will be weaker when bumps are initiated in two different layers. Also, note in the collocation events discussed above, bumps are attracted to the same relative position within a layer, but two bumps still remain. As before, we consider a task wherein the target $\phi_1$ is probed at the end of a delay-period and compute the MSE according to Eq.~(\ref{msedef}).

One distinction in comparison to the single layer network case is that bumps that begin close enough, $\phi_2 - \phi_1 < 2 \Delta_{x12}^c = h/(1-\e^{-2h/\sigma_{jk}})$ (Fig. \ref{fig12_multilay}C), will collocate and limit each others' diffusion via the attraction of interlaminar connections. The stochastic dynamics of this effect has been analyzed in detail in \cite{kilpatrick13c,bressloff15} assuming the relative distance between bumps is small. In essence, a linear approximation is applied to Eq.~(\ref{laycent}) and the variance of $(\Delta_1,\Delta_2)$ can be computed directly for the resulting multivariate Ornstein-Uhlenback process
\begin{subequations} \label{lindublay}
\begin{align}
\d \Delta_1 &= \kappa \left[ -\Delta_1 + \Delta_2 \right] \d t + \d \bZ_1, \\
\d \Delta_2 &= \kappa  \left[ +\Delta_1 - \Delta_2 \right] \d t + \d \bZ_2,
\end{align}
\end{subequations}
where $\kappa = (w_{c}(0)-w_{c}(2h))/\bar{\alpha}$ ($w_c(x):=w_{12}(x)=w_{21}(x)$) and
\[
\d \bZ = \frac{\sqrt{\ep \cdot \theta}}{2 \bar{\alpha}} \left( \begin{array}{c} \d {\mc Z}_1(\Delta_1,t) \\  \d {\mc Z}_2(\Delta_2,t) \end{array} \right),
\]
so that for symmetric and independent noise, $\langle \bZ_1(t) \bZ_2(t) \rangle \equiv  0$ and $\langle \bZ_1^2(t) \rangle = \langle \bZ_2^2(t) \rangle  = D t$ with $D$ defined as in Eq.~(\ref{diffcon}). We can in fact directly compute the first and second moments of $\Delta_1$ from Eq.~(\ref{lindublay}) as in \cite{kilpatrick13c}. For simplicity, we present our results for the case of symmetric initial conditions $\phi_1 = - \phi_2 = x_0$:
\begin{align*}
\langle \Delta_1(t) \rangle &= x_0 \e^{-2 \kappa t}, \\
\langle \Delta_1^2(t) \rangle & = x_0^2 \e^{-4 \kappa t} +  \frac{D}{2} t + \frac{D}{8\kappa} \left[ 1 - \e^{-4\kappa t} \right] .
\end{align*}
Thus, for collocating bumps, we can approximate Eq.~(\ref{msedef}), similar to Eq.~(\ref{mergmse}) as
\begin{align*}
{\mc M}_{\rm mg}& = \langle (\Delta_{1-\text{out}} - \phi_1)^2 \rangle \\
&= \langle \Delta_{1-\text{out}}^2 \rangle - 2 x_0 \langle \Delta_{1-\text{out}} \rangle + x_0^2  \\
 &= \frac{D}{2}T + \frac{D}{8\kappa} \left[ 1 - \e^{-4\kappa T} \right] +x_0^2 \left( 1 - \e^{-2 \kappa T} \right)^2,  
\end{align*}
so notice that the interlaminar coupling dampens the long-term fluctuations in the bumps' positions so the diffusion is half that of the single-layer network~\citep{kilpatrick13c}. Additional contributions arise from the competition between stochastic fluctuations driving the bumps apart and the coupling pulling them together. This gives us an analytic approximation of the MSE for collocating bumps, $x_0 < \Delta_{x12}^c$.

When bumps do not attract one another, then we can use our nonlinear approximation Eq.~(\ref{laycent}) to estimate the error in recall via direct simulation. We compare this approximation to simulations of the multilayered neural field, Eq.~(\ref{mfield}) in Fig. \ref{fig12_multilay}E. The distance-dependence of MSE is demonstrated in comparison to the MSE in the single layer neural field, Eq.~(\ref{nfield}). As expected, because the connectivity between layers is weaker and narrower than connectivity within layers, bumps that are initiated in the same layer interact more strongly than bumps in different layers. For example, items of the same color will tend to interact in WM more strongly than items of different colors. We expect this result will extend to the case of randomized item locations, and will explore this more in subsequent work.

We conclude that an interacting bumps model of multi-item WM can capture several key features of error. Synaptic fluctuations lead to the time-dependent scaling typically observed in parametric WM tasks~\citep{white94,ploner98,wimmer14}. More important for multi-item WM tasks is the impact of item number on the reliability of storage~\citep{ma14}. Here, we have shown that interactions in item memory can be described by the dynamics of multiple bumps in a common network. Specifically, items that are closer to one another in one or more feature dimensions will have associated bumps that interact more strongly, potentially leading to bump annihilations or repulsions. Notably, our mechanism is much more suggestive of the resource model of WM~\citep{bays08} than of a slots model, which would assume that the first few items stored do not have associated memories that interact in any way~\citep{zhang08}.

\section{Discussion}
\label{discussion}

Working memory is a central feature of cognition, which plays an important role in attention~\citep{gazzaley12} and motor planning~\citep{ikkai11}. Limitations on the fidelity of working memory can therefore limit other cognitive functions. We have proposed a simplified model to account for item and temporal limitations in multi-item WM, based on recent studies of spiking networks~\citep{wei12,almeida15}. Similar to these previous studies, we associate the memory of an item in space with the location of a bump attractor, subject to fluctuations and interactions with other bumps.

The advantage of our model is that we were able to analyze the dynamics of the network, and reduce the dynamics to a low-dimensional system describing the interfaces of the bumps. Errors in recall occur due to merging, repulsion, and annihilation events resulting from bump interactions, which can also be captured by the corresponding interface equations. This is in addition to the typical fluctuation-induced errors known to arise during the delay-period of visual WM tasks~\citep{wimmer14,constantinidis16}. Importantly, we have shown that the strength of synaptic coupling in a recurrent network shapes the mean squared error (MSE) in WM tasks. More weakly coupled networks support narrower bump attractors that interact less, but are more subject to noise fluctuations. Strongly coupled networks possess wider bumps, which are less subject to noise fluctuations. There is an optimal scaling of synaptic strength that minimizes the MSE, trading off reduced effects of noise with bumps that do not interact too strongly. This optimal scaling strength decreases for tasks requiring memory of more items $N$.

Interface methods were used to reduce the neural field model to a system of a few differential equations, corresponding to the threshold crossing points of dynamically evolving bumps. An exact description of the evolution of the interfaces can be obtained by evolving an integral equation describing the dynamically evolving gradient at the interfaces~\citep{coombes12}. However, these integral equations are much less straightforward to derive for the case in which stochastic forcing is incorporated into the evolution equations, so we employed a static gradient approximation, ignoring the perturbations of the gradient near the interfaces. We expect that interface equations that capture these fluctuations in the gradient would provide a more accurate approximation of the dynamics of the full neural field model, Eq.~(\ref{nfield}). Nonetheless, we were able to derive a reasonably accurate approximation for the dynamics of multiple remembered items during the delay period of a WM task. This low-dimensional description could be leveraged in future work to explore the effects of bump interactions in two- and three-dimensional feature space.

Controversy still remains as to whether errors in multiple item WM are best explained using a slots~\citep{zhang08} or resource~\citep{bays08} model of item storage. Our model mostly supports the latter hypothesis, since recall errors depend on the item number across all item counts. Furthermore, we expect that the practical capacity of the network will be relatively high, since many bumps can be stored in a single recurrent network, and there will not tend to be an abrupt drop in accuracy at any particular item count. Thus, we see no strong evidence of a small and fixed capacity in our model. Rather, the capacity of the network is much larger, and would likely not be revealed by the typical single digit item counts used in WM experiments~\citep{ma14}. Furthermore, our model predicts that if two angles are initially placed close to one another, there will be more error in their recalled locations. This is due to the nonlinear interactions in the network. This is consistent with a recent experimental study of multi-item working memory tasks analyzed in \cite{almeida15}.

Computational models of multi-item WM that use multiple activity bumps have been explored previously in the works of \cite{macoveanu06,edin09,wei12}. However, none of these works provided a concise description of the neural and synaptic mechanics underlying specific forms of recall error, as we have done using our interface equations approach. Primarily, these previous works explored trends in large-scale simulations. One exception is \cite{edin09}, who were able to derive a steady-state estimate of the capacity of their network, showing it to be somewhere between two and seven items due to inhibition decreasing the excitability of inactive regions of the network. Our network does not exhibit this low capacity, likely due to the fact that excitation and inhibition are balanced so the total excitatory and inhibitory input to the network sums to zero. This approach is motivated by recent work suggesting the common features of cortical variability arise from balanced and spatially-organized excitation and inhibition~\citep{lim14,rosenbaum17}. In bump attractor models of WM, we expect the mean network activity will tend to increase with item number in networks imbalanced towards excitation~\citep{edin09}, and stay relatively constant in networks with roughly balanced excitation/inhibition~\citep{wei12}. In this regard, our results are more in line with \cite{wei12}, especially because \cite{edin09} consider the incorporation of external inputs which further modifies the network's mean excitability.

We could also extend our analyses to consider WM for objects residing in a higher-dimensional feature space. For instance, we could carry out a more thorough analysis of the mulitlayer neural field model introduced in Section \ref{multfeat}, or analyze the problem of a stimulus space covering two continuous dimensions. Note that working memory tasks usually require subjects to remember both the color and position of remembered items, so that, for example, the color can be used to indicate which item the subject should recall~\citep{bays08}. Studying the interactions of neural activity along two stimulus dimensions would also aid in providing a mechanistic explanation for swap errors. Swap errors can occur when a subject uses the stored location of a different item to report the remembered location of a cued item~\citep{bays09}. Thus, it may be that both item interactions and fluctuations that produce errors in the stored position may affect the stored color in a similar way. Recently, \cite{schneegans17} considered a related neural population model that could simultaneously represent color, orientation, and location of a stimulus, an extension of a model initially discussed in \cite{bays14}. Indeed, stochasticity in the neural code could account for swap errors and response variability, although they did not consider the detailed dynamics of the neural populations during a delay period. 

Our mechanistic model of multiple item WM limitations has a distinct advantage over heuristic parameterized models, in that it is linked to constraints of physiology. However, we expect there are many extensions of the framework, which would likely more reliably reflect actual physiology. In particular, while there is evidence for persistent neural activity in visual WM, we expect that the strongly bistable nature of bumps in Eq.~(\ref{nfield}) may be inconsistent with the various accumulating and decaying activity traces observed in some cortical during WM tasks~\citep{murray16,zylberberg17}. Such temporally heterogeneous activity, distinct from the relatively stable trace a bump attractor, may still provide a stable population code. Thus, it would be interesting to explore how dynamic neural activity traces representing different items would interact in a computational model, and whether the principles of a resource-type model of WM would arise there as well.

\section*{Acknowledgements}

NK was supported by EXTREEMS - QED: Directions in Data Discovery in Undergraduate Education (NSF DMS-1407340). ZPK was supported by an NSF grant (DMS-1615737).

\bibliographystyle{spbasic} 
\bibliography{interacting}
     % basic style, author-year citations

\end{document}